\documentclass{article}
\usepackage{fullpage}
\usepackage{amsmath}           		          	
\usepackage{graphicx,epstopdf}					
\usepackage{amssymb}
\newcommand {\bxi}{\mbox{\boldmath$\xi$}}
\pdfoutput = 1 
\allowdisplaybreaks

\begin{document}
\title{\bf Response of a magnetically diverted tokamak plasma to a resonant magnetic perturbation}
\author{Richard Fitzpatrick\,\footnote{rfitzp@utexas.edu}\\[1ex]{\em Institute for Fusion Studies, Department of Physics,}\\[0.5ex] {\em University of Texas at Austin, Austin TX 78712, USA}}
\date{}

\maketitle

\begin{abstract}
The safety-factor profile of a magnetically diverted tokamak plasma diverges logarithmically as the magnetic separatrix 
(a.k.a.\ the last closed magnetic flux-surface) is approached. At first sight, this suggests that, when determining the response of such a plasma to a static,  externally generated,  resonant
magnetic perturbation (RMP), it is necessary to include an infinite number of rational magnetic flux-surfaces (at which the RMP resonates with the equilibrium magnetic field) in the calculation, the
majority of which lie  very close to the separatrix. In fact, when finite plasma resistivity is taken into account, this turns out not to be the case. 
Instead, it is only necessary to include rational surfaces that
lie in the region $0<{\mit\Psi}<1-\epsilon_c$, where ${\mit\Psi}$ is the normalized poloidal magnetic flux, and
$0<\epsilon_c\ll 1$. The parameter $\epsilon_c$ can be calculated from the edge plasma parameters. It is estimated that $\epsilon_c=1.5\times 10^{-3}$ for an $n=1$
RMP, and $\epsilon_c=4.8\times 10^{-3}$ for an $n=4$ RMP,  in  a typical JET H-mode plasma. 
\end{abstract}

\section{Introduction}
All modern tokamak plasmas feature a {\em magnetic divertor},  which is a particular configuration of the magnetic field that redirects the outermost field-lines  away from the confined plasma region and into a controlled exhaust area where the plasma particle and heat fluxes can be safely  absorbed by a solid  target plate \cite{wesson}. Each magnetic flux-surface in a tokamak plasma has an associated value of the {\em safety-factor}, which specifies the  number of
toroidal circuits of the surface that a constituent  magnetic field-line makes per poloidal circuit \cite{meiss,boozer}. Magnetic flux-surfaces possessing rational
values of the safety-factor are special. Indeed, externally generated, static, non-axisymmetric magnetic perturbations---known as {\em resonant magnetic perturbations} (RMPs)---can drive  magnetic reconnection  on such {\em rational surfaces}, leading to localized changes in magnetic
topology that degrade the ability of the surrounding flux-surfaces to confine energy \cite{chang}. Alternatively, shielding currents can flow, predominately parallel 
to magnetic field-lines, on rational surfaces, and act to suppress driven  magnetic
reconnection \cite{rf1}. In general, neither driven magnetic reconnection nor shielding currents can occur on irrational magnetic flux-surfaces.

A magnetically diverted tokamak plasma possesses a so-called ``last closed magnetic flux-surface" (LCFS). Magnetic flux-surfaces that lie inside the
LCFS are topologically simple tori, and are completely occupied by plasma. Magnetic flux-surfaces that lie outside the LCFS have more complicated
topology, and are only partially occupied by plasma. The LCFS features a magnetic ``X-point", which is a circular loop, coaxial with the plasma torus, on
which the poloidal magnetic field is zero. (In some very  special tokamak plasmas, the LCFS features two X-points. However, we shall 
not consider such plasmas here.) As the LCFS is approached, the safety-factor exhibits a logarithmic singularity \cite{pom}, as a direct consequence of the presence of the X-point, and there is an
associated accumulation of rational magnetic flux-surfaces in the vicinity of the LCFS. In fact, for an RMP with a given toroidal mode number, there are,
in principle, an {\em infinite}\/ number of associated rational surfaces, most of which lie  very close to the LCFS,  on which the RMP can drive magnetic reconnection. It is
impossible to include all of these surfaces in a practical plasma response calculation. Hence, the standard approach is to simply ignore rational flux-surfaces associated with very high values of the
safety-factor \cite{mars,ipec,dcon,park,epec,gpec,logan}. The purpose of this paper is to investigate to what extent this approach is justified. 

This paper is organized as follows. A particularly simple model of a magnetically diverted tokamak plasma is introduced in Sect.~\ref{stwo}, and is
used to define exactly what is meant by a straight-field-line  coordinate system, the safety-factor profile, and a rational surface, both inside and outside the LCFS. 
A more realistic model of a magnetically diverted tokamak plasma is presented in Sect.~\ref{sthree}. The response of a magnetically diverted plasma
to an RMP is analyzed  in Sect.~\ref{response}. Section~\ref{separatrix}  investigates the predictions of the plasma
response model in the vicinity of the LCFS. Finally, the paper is summarized in Sect.~\ref{sum}. 

\section{Simple model of a magnetically diverted plasma}\label{stwo}
\subsection{Introduction}
The aim of this section is to construct a very simple model of a magnetically diverted tokamak plasma.

\subsection{Equilibrium magnetic field}
Let $x$, $y$, $z$ be conventional Cartesian coordinates. Suppose that two current filaments run parallel to the $z$-axis \cite{pom,reiman}. Let the first filament carry the
current $I_p$, and pierce the $x$-$y$ plane at $x=y=0$. Let the second filament carry the current $I_c$, and
pierce the $x$-$y$ plane at $x=0$, $y=-a$.  The first filament represents the ``toroidal'' (i.e., $z$-directed) plasma
current, whereas the second represents the toroidal current flowing in the magnetic divertor coil that generates the magnetic X-point. 
Suppose that there is a uniform, externally generated, toroidal magnetic
field of strength $B_0$. Let the system be periodic in the $z$-direction with period $2\pi\,R_0$, where $R_0$ is the simulated major
radius of the plasma.  
It is helpful to define the simulated toroidal angle,  $\phi= z/R_0$. 

The equilibrium  magnetic field can be written in the divergence-free manner
\begin{equation}\label{e1}
{\bf B} = \nabla\phi\times \nabla\psi_p + B_0\,R_0\,\nabla\phi,
\end{equation}
where 
\begin{equation}
\psi_p(x,y)=  \frac{\mu_0\,I_p\,R_0}{4\pi}\,\ln\left(x^2+y^2\right) + \frac{\mu_0\,I_c\,R_0}{4\pi} \,\ln\left[x^2+(y+a)^2\right]
\end{equation}
is the ``poloidal'' (i.e., circulating in  the $x$-$y$ plane) magnetic flux (divided by $2\pi$) generated by the two current filaments.

\subsection{Straight-field-line coordinates}
It is convenient to re-express the magnetic field in the standard Clebsch form
\begin{equation}\label{e2}
{\bf B} = \nabla(\phi-q\,\theta)\times \nabla \psi_p,
\end{equation}
where $\theta$ is a poloidal angle, and  $q=q(\psi_p)$ the (dimensionless) {\em safety-factor}  \cite{boz}. Equations~(\ref{e1}) and (\ref{e2}) can be reconciled provided 
\begin{equation}\label{e3}
\nabla\psi_p\times\nabla\theta\cdot\nabla\phi = \frac{B_0}{R_0\,q}.
\end{equation}
Note, from Eq.~(\ref{e2}),  that ${\bf B}\cdot\nabla\psi_p=0$, which implies that  $\psi_p$ is a magnetic flux-surface label. Furthermore,
${\bf B}\cdot\nabla(\phi-q\,\theta)=0$, which implies that magnetic field-lines within a given flux-surface appear as straight lines, 
with gradient $d\phi/d\theta = q$, when 
plotted in the $\theta$-$\phi$ plane. In fact, $\psi_p$, $\theta$, $\phi$ are known as {\em straight-field-line coordinates}, and $\theta$
is termed a {\em straight}\/  poloidal angle \cite{boz}. The defining property of a straight-field-line coordinate system is the choice $q=q(\psi_p)$,
rather than $q=q(\psi_p,\theta)$, or even $q=q(\psi_p,\theta,\phi)$. There are many types of straight-field-line coordinate
systems (e.g., Hamada \cite{hamada}, Boozer \cite{bcoord}, PEST \cite{pest}, equal-arc), all of which have the required property that $q=q(\psi_p)$. For the sake of convenience, in this paper we shall adopt PEST
coordinates, which are characterized by a toroidal angle, $\phi$, that corresponds to the angular coordinate in the $R$, $\phi$, $Z$ cylindrical 
coordinate system (that is coaxial with the plasma torus). (Here, we are looking ahead to the generalization to
true toroidal geometry that takes place in Sect.~\ref{equil}.) Once the choice of toroidal angle is made, the coordinate system is uniquely defined. 

\subsection{Non-diverted edge safety-factor}
In the absence of the divertor current, the plasma would have a circular cross-section in the $x$-$y$ plane of minor radius $a$, and
an edge safety-factor value of
\begin{equation}\label{e4}
q_\ast = \frac{2\pi\,B_0\,a^2}{\mu_0\,I_p\,R_0}.
\end{equation}

\subsection{Normalization scheme}
Let  $x=a\,X$, $y=a\,Y$,  $\nabla = a^{-1}\,\hat{\nabla}$, and $\psi_p = \mu_0\,I_p\,R_0\,\psi/(2\pi)$.
It follows that
\begin{align}
\hat{\nabla}\psi\times \hat{\nabla}\theta\cdot\hat{\nabla}\phi &= \frac{a}{R_0}\,\frac{q_\ast}{q},\label{ex}\\[0.5ex]
\psi(X,Y) &= \frac{1}{2}\,\ln\left(X^2+Y^2\right) + \frac{\zeta}{2}\,\ln\left[X^2+(Y+1)^2\right],\\[0.5ex]
\psi_X(X,Y) &= \frac{X}{X^2+Y^2} + \frac{\zeta\,X}{X^2+(Y+1)^2},\\[0.5ex]
\psi_Y(X,Y) &= \frac{Y}{X^2+Y^2} +\frac{\zeta\,(Y+1)}{X^2+(Y+1)^2},
\end{align}
where $\zeta=I_c/I_p$. Here, $\psi_X\equiv\partial\psi/\partial X$, et cetera. 

\subsection{Magnetic X-point}
The magnetic $X$-point is located at the point in the $X$-$Y$ plane where  $\psi_X=\psi_Y=0$ (i.e., where the poloidal magnetic
field-strength is zero).
As is easily demonstrated, the coordinates of this point are ($X_x$, $Y_x$), where  $X_x=0$ and $Y_x=-1/(1+\zeta)$. 
The so-called magnetic separatrix corresponds to the curve $\psi(X,Y)= \psi_x$, where 
\begin{equation}
\psi_x \equiv \psi(X_x,Y_x) = \ln\left[\frac{\zeta^{\,\zeta}}{(1+\zeta)^{1+\zeta}}\right].
\end{equation}
 It is helpful to define the normalized poloidal flux, ${\mit\Psi}= \psi_x/\psi$. (Note that this definition is different from the
 conventional one, ${\mit\Psi}= \psi/\psi_x$, because $\psi\rightarrow \infty$ on the magnetic axis, $X=Y=0$, and
 we wish the normalized flux to increase as we go from the axis to the separatrix.)

\subsection{Construction of straight-field-line coordinate system}
Equation~(\ref{ex})  yields
\begin{equation}
\frac{d\theta}{dL} = \frac{q_\ast}{q\,|\hat{\nabla}\psi|}.
\end{equation}
where $dL$ is an element of normalized length (in the $X$-$Y$ plane) around a magnetic flux-surface. 
It follows that
\begin{equation}
q(\psi) = \frac{q_\ast}{2\pi}\oint\frac{dL}{|\hat{\nabla}\psi|},
\end{equation}
where $\oint$ implies a complete circuit in $\theta$ at constant $\psi$. 
It is easily demonstrated that, on such a circuit, 
\begin{align}\label{e11}
\frac{dX}{dL} &= -\frac{\psi_Y}{\sqrt{\psi_X^{\,2} + \psi_Y^{\,2}}},\\[0.5ex]
\frac{dY}{dL}&= \frac{\psi_X}{\sqrt{\psi_X^{\,2} + \psi_Y^{\,2}}},\\[0.5ex]
\frac{d\phi}{dL} &=  \frac{q_\ast}{\sqrt{\psi_X^{\,2} + \psi_Y^{\,2}}},\\[0.5ex]
\frac{d\varpi}{dL}  &= \frac{X\,\psi_X + Y\,\psi_Y}{(X^2+Y^2)\sqrt{\psi_X^{\,2} + \psi_Y^{\,2}}},\label{e22}
\end{align}
where
\begin{equation}
\varpi =\tan^{-1} \left(\frac{Y}{X}\right)
\end{equation}
is a geometric poloidal angle. 
Here, $\phi$ is calculated on the assumption that we are following a magnetic field-line within the flux-surface (i.e., $d\phi/d\theta = q$). We need to
integrate Eqs.~(\ref{e11})--(\ref{e22}) from $\varpi=0$ to $\varpi = 2\pi$, subject to the initial
condition $\phi(\varpi=0)=0$, and then set $q(\psi)= \phi(\varpi=2\pi)/(2\pi)$.
We can then compute $\theta$ using
\begin{align}
\frac{d\theta}{dL}& = \frac{q_\ast}{q\sqrt{\psi_X^{\,2} + \psi_Y^{\,2}}}.
\end{align}

\begin{figure}[t]
\centerline{\includegraphics[width=\textwidth]{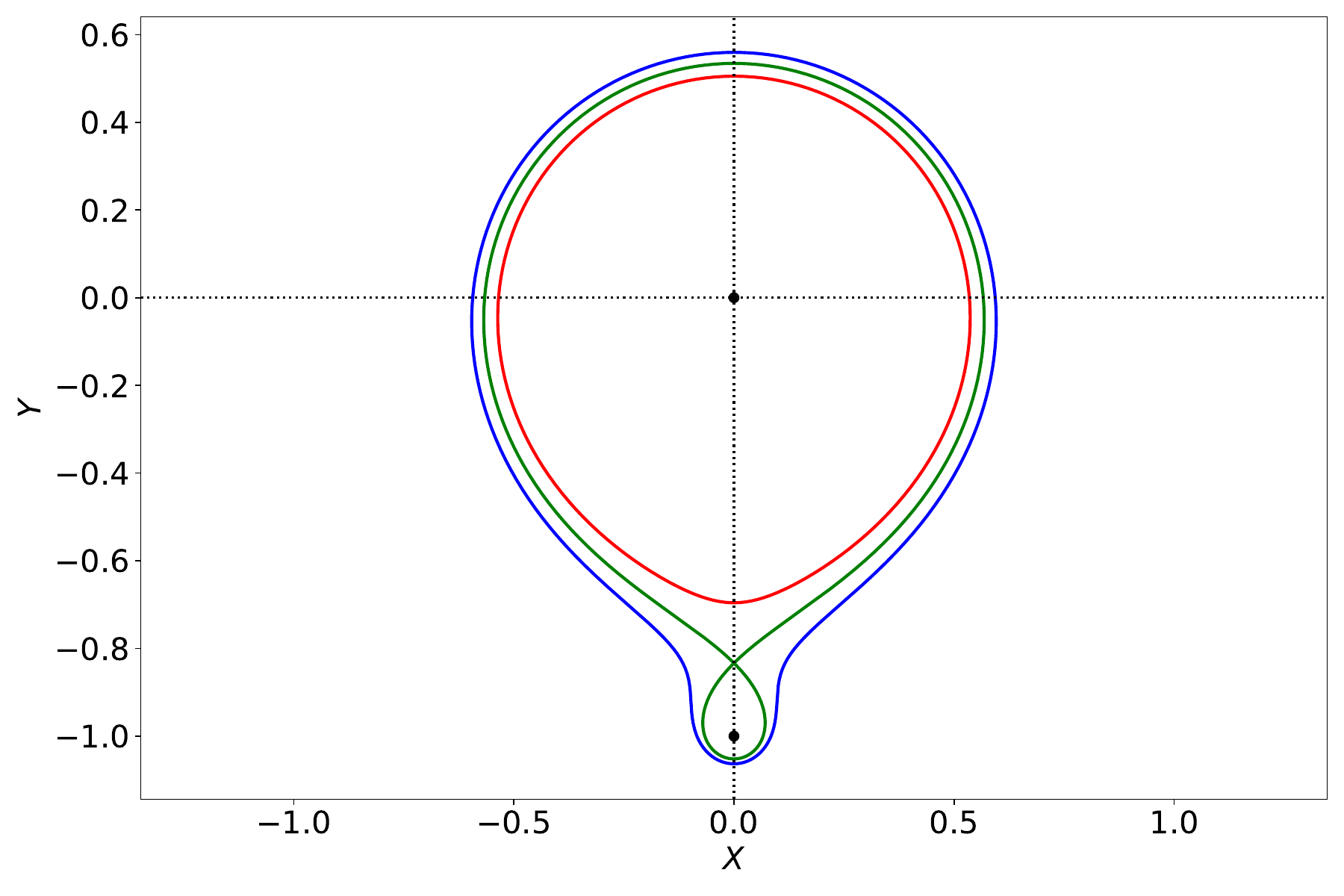}}
\caption{The magnetic flux-surfaces ${\mit\Psi}=0.9$ (red), ${\mit\Psi}=1.0$ (green), and ${\mit\Psi}=1.1$ (blue) in the absence of a divertor plate. The black dots shows the locations of the two current filaments. Here, $q_\ast=12$ and $\zeta=0.2$. }\label{fig1}
\end{figure}

Let $q_\ast = 12$ and $\zeta=0.2$. 
Figure~\ref{fig1} shows the magnetic flux-surfaces ${\mit\Psi}=0.9$, ${\mit\Psi}=1.0$, and ${\mit\Psi}=1.1$, plotted in the 
$X$-$Y$ plane. Flux-surfaces characterized by ${\mit\Psi}<1$ do not enclose the divertor coil filament, whereas those
characterized by ${\mit\Psi}>1$ do enclose the filament. The {\em magnetic separatrix}, ${\mit\Psi}=1$, separates flux-surfaces
 that do and do not enclose the divertor coil filament, and crosses itself at the magnetic X-point. 
 
 \begin{figure}[t]
\centerline{\includegraphics[width=\textwidth]{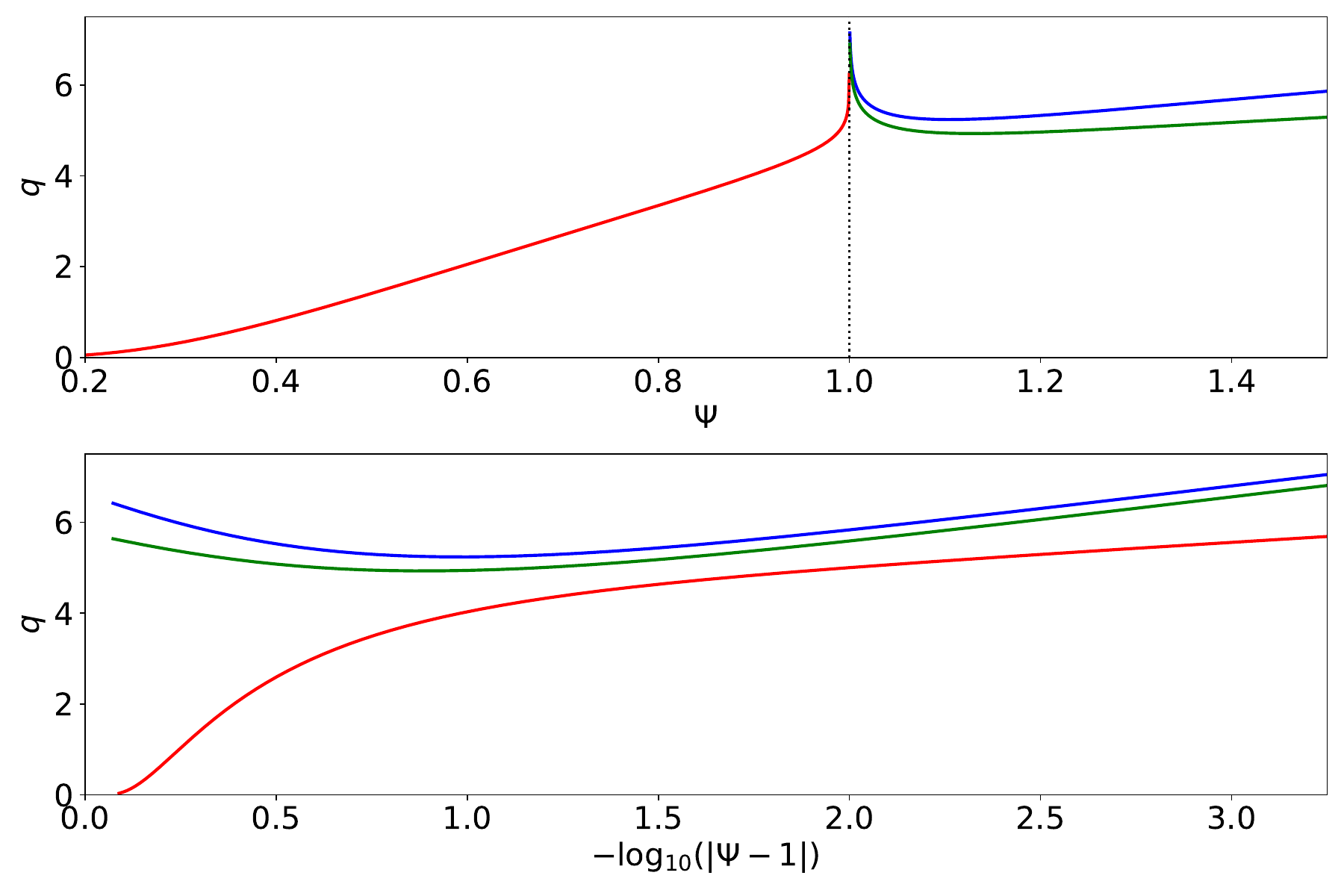}}
\caption{The safety-factor profile, $q({\mit\Psi})$. The red curve shows the safety-factor
inside the magnetic separatrix. The blue and green curves show the safety-factor outside the separatrix in the absence and
in the presence of a divertor plate, respectively.  Here, $q_\ast=12$ and $\zeta=0.2$. }\label{fig2}
\end{figure}

The red and  blue curves in Fig.~\ref{fig2} show the safety-factor profile, $q({\mit\Psi})$, inside and outside the magnetic separatrix, respectively. It is clear that the safety-factor generally increases with increasing ${\mit\Psi}$. 
However,   $q\rightarrow\infty$ as ${\mit\Psi}\rightarrow 1$.
 In other words, the safety-factor tends to infinity as the magnetic separatrix is approached from either direction. It is apparent from the bottom
 panel of Fig.~\ref{fig2} that $q$ approaches infinity {\em logarithmically}\/ as ${\mit\Psi}\rightarrow 1$ (because the plot of $q$ versus
 $\log_{10}(|{\mit\Psi}-1|)$ asymptotes to a straight line as $|{\mit\Psi}-1|\rightarrow 0$). In other words, close to
 the separatrix,  we can write
 \begin{equation}
 q({\mit\Psi})\simeq -\alpha_-\,\ln(1-{\mit\Psi})
 \end{equation}
 for ${\mit\Psi}<1$, and 
 \begin{equation}
 q({\mit\Psi})\simeq -\alpha_+\,\ln({\mit\Psi}-1)
 \end{equation}
 for ${\mit\Psi}>1$. 
 Moreover, the figure implies that $\alpha_+>\alpha_-$. 
 
\begin{figure}[t]
\centerline{\includegraphics[width=\textwidth]{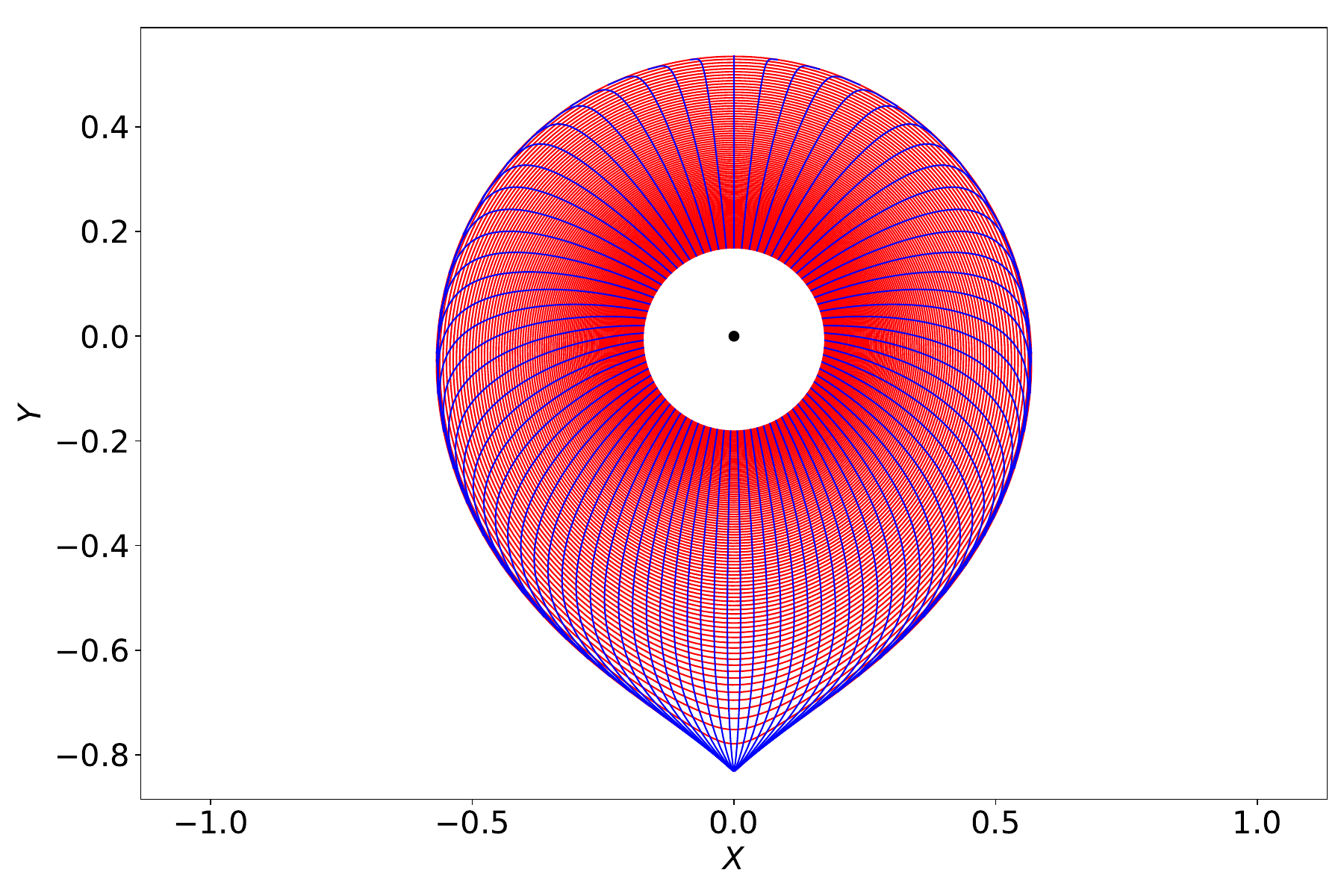}}
\caption{The straight-field-line coordinate system inside the magnetic separatrix. The red curves
are surfaces of constant $\hat{\psi}$, whereas the blue curves are surfaces of constant $\theta$. The black dot shows the location of the plasma current filament. Here, $q_\ast=12$ and $\zeta=0.2$.}\label{fig3}
\end{figure}

\begin{figure}[t]
\centerline{\includegraphics[width=\textwidth]{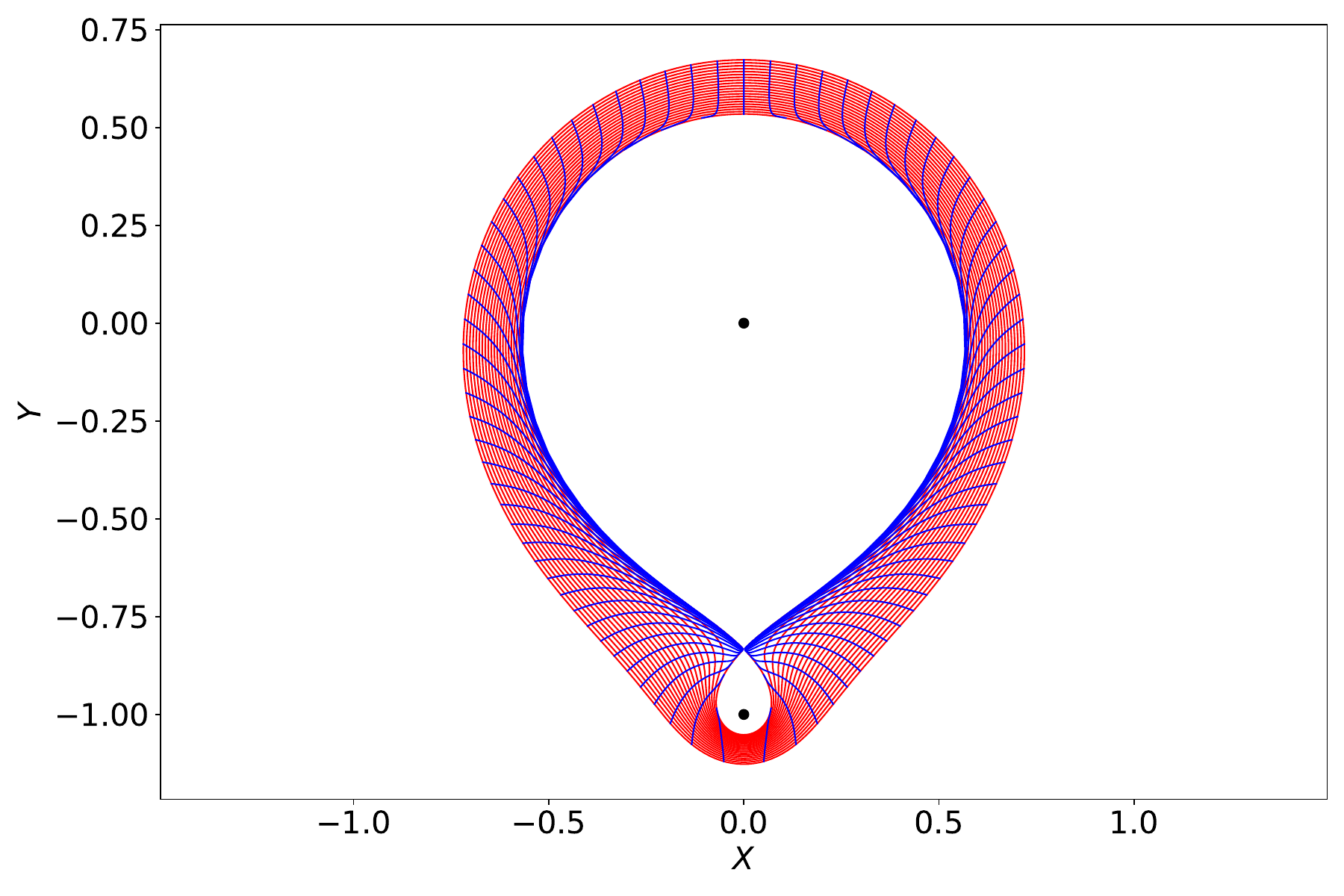}}
\caption{The straight-field-line coordinate system outside the magnetic separatrix in the absence of a divertor plate. The red curves
are surfaces of constant $\hat{\psi}$, whereas the blue curves are surfaces of constant $\theta$. The black dots shows the locations of the two current filaments.  Here, $q_\ast=12$ and $\zeta=0.2$.  }\label{fig4}
\end{figure}

 Figures~\ref{fig3} and \ref{fig4} show the straight-field-line coordinate system inside and outside the magnetic separatrix, respectively. It can be seen that, as the magnetic
 separatrix is crossed, all of the contours of $\theta$ converge onto, and then diverge away from, the X-point. This singular behavior occurs
 because $|\hat{\nabla}\psi|\rightarrow \ell$ and $|\hat{\nabla}\theta|\rightarrow 1/[\ell\,\ln|{\mit\Psi}-1|]$ as the X-point is approached, where $\ell$ represents 
 distance from the X-point in the $X$-$Y$ plane
 
\subsection{Significance of straight-field-line coordinates}
To understand the significance of the straight-field-line coordinate system, suppose that the plasma is subject to a static (in the laboratory frame) magnetic perturbation that
varies with $\theta$ and $\phi$ as  $\exp[\,{\rm i}\,(m\,\theta-n\,\phi )]$. Here, $m$ and $n$ are integers. 
In other words, the perturbation  (which is, of course, single-valued in the angular coordinates $\theta$ and $\phi$) possesses
$m$ periods in the poloidal angle,  and $n$ periods in the toroidal angle.  The curl of the perturbed, linearized,
electron fluid equation of motion yields
\begin{equation}\label{e21}
 n\,(\omega_E+\omega_{\ast\,e})\left[b^{\psi_p} -\frac{B_0}{R_0}\left(\frac{m}{q}-n\right)\xi^{\psi_p}\right] \simeq \frac{B_0}{R_0}\,\frac{m}{q} \,\eta_\parallel \,j_\phi,
\end{equation}
where subscripts and superscripts denote covariant and contravariant components in the $\psi_p$, $\theta$, $\phi$ coordinate
system, respectively \cite{meiss}.  Here, ${\bf b}$ is the perturbed magnetic field, $\bxi$ the Lagrangian electron fluid displacement,  ${\bf j}$ the perturbed current
density, and $\eta_\parallel$ the plasma parallel electrical resistivity. Moreover,
\begin{align}
\omega_E(\psi_p) = -\frac{d{\mit\Phi}}{d\psi_p}
\end{align}
is the {\em E-cross-B frequency}, whereas
\begin{align}
\omega_{\ast\,e}(\psi_p) = \frac{1}{e\,n_e}\,\frac{dp_e}{d\psi_p}
\end{align}
is the {\em electron diamagnetic frequency}. 
Here, ${\mit\Phi}(\psi_p)$ is the electrostatic potential,  $p_e(\psi_p)$  the electron pressure, $n_e(\psi_p)$ the electron number density, and $e$ the magnitude of the electron charge. 

Equation~(\ref{e21}) describes how  
the inductive electric
field generated by a time-varying magnetic field, as seen in rest frame of the local electron fluid, attempts to drive  a current that runs parallel to magnetic field-lines. On a general magnetic flux-surface, the
two terms (in the large square brackets) on the left-hand side of the equation cancel one another out, and there is no  driven current. In other words, 
the electron fluid displaces rather than allowing a parallel inductive current to flow \cite{rf1}. However, it is clear from the equation that there exists 
a special magnetic flux-surface, termed  a {\em rational}\/ flux-surface, at which the safety-factor takes the rational
value $q=m/n$ \cite{boz}. On the rational flux-surface, the two terms on the left-hand side of Eq.~(\ref{e21}) cannot cancel one another
out, because the second term is zero everywhere on the surface. Hence, in general, a parallel inductive current is driven on the rational flux-surface. The current is a
{\em shielding current}\/ that acts to suppress driven magnetic reconnection on the flux-surface \cite{rf1,rf2}. 

We can now appreciate that, by employing a straight-field-line coordinate system, we can distinguish rational magnetic flux-surfaces from
irrational flux-surfaces. [A rational surface is one on which the associated safety-factor can be expressed as a rational number. Note that
this definition only makes sense if the safety-factor is a flux-surface function. In this respect, we cannot agree with the analysis of Ref.~\cite{lj} where a
coordinate system is adopted in which $q=q(\psi_p,\theta)$ in the vicinity of the X-point.] We can also determine the angular variation of the particular magnetic perturbation that drives a
shielding current on a particular rational surface. Finally, it is clear from  that Figs.~\ref{fig2}--\ref{fig4}  that rational
flux-surfaces exist both inside and outside the magnetic separatrix. 

\subsection{Effect of divertor plate}
Rational magnetic flux-surfaces exist outside the magnetic separatrix  \cite{reiman} because flux-surfaces in this region form closed loops, that complete below the
divertor coil filament, when plotted in
the $X$-$Y$ plane. In fact, this must be the case in an axisymmetric system otherwise the divergence of the magnetic field would be non-zero.
Thus, the commonly used term  ``last closed magnetic flux-surface" (meaning the magnetic separatrix) is inaccurate. All of the magnetic
flux-surfaces are closed. However, those outside the magnetic sepatatrix are only partially 
occupied by plasma because they intersect the solid divertor plate.

 Suppose that
the divertor plate is horizontal (i.e., parallel to the $X$-axis), and is situated halfway between the X-point and the divertor coil filament. 
When integrating Eqs.~(\ref{e11})--(\ref{e22}), we are really integrating along the path of a shielding current filament excited by an inductive electric field.
 Under normal circumstances, the  filament is constrained to run parallel to magnetic
field-lines. However, within the rigid divertor plate (which is assumed to be electrically conducting), the filament  takes the path of least electrical resistance, which implies that it runs along the
$X$-axis at constant $\phi$. It follows that, within the divertor plate, Eqs.~(\ref{e11})--(\ref{e22}) must be replaced by
\begin{align}
\frac{dX}{dL} &= 1,\\[0.5ex]
\frac{dY}{dL}&= 0,\\[0.5ex]
\frac{d\phi}{dL} &=  0,\\[0.5ex]
\frac{d\varpi}{dL}  &=-\frac{Y}{X^2+Y^2}. 
\end{align}

\begin{figure}[t]
\centerline{\includegraphics[width=\textwidth]{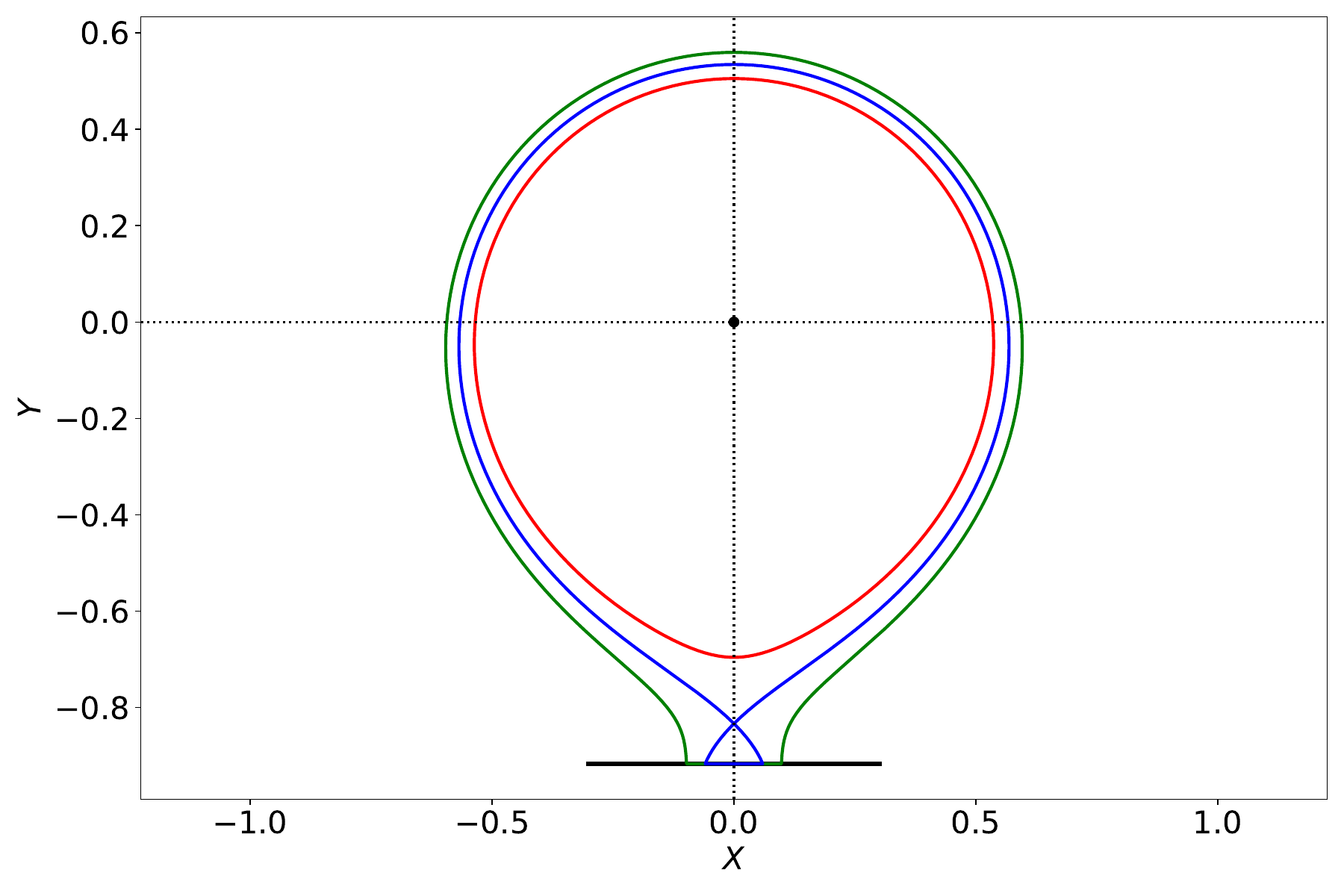}}
\caption{The magnetic flux-surfaces ${\mit\Psi}=0.9$ (red), ${\mit\Psi}=1.0$ (green), and ${\mit\Psi}=1.1$ (blue)  in the presence of a divertor plate located halfway between the X-point and the divertor coil filament. The black dot shows the location of plasma current filament, and the horizontal black line shows the location of the divertor plate. Here, $q_\ast=12$ amd $\zeta=0.2$. }\label{fig5}
\end{figure}

Again, let $q_\ast = 12$ and $\zeta=0.2$. 
Figure~\ref{fig5} shows the magnetic flux-surfaces ${\mit\Psi}=0.9$, ${\mit\Psi}=1.0$, and ${\mit\Psi}=1.1$, plotted in the 
$X$-$Y$ plane. Note that the surfaces ${\mit\Psi}=1.0$ and ${\mit\Psi}=1.1$ both have sections that run parallel to
the divertor plate. Strictly speaking, these sections are not magnetic flux-surfaces (the actual magnetic flux-surfaces complete below the divertor plate, as shown in Fig.~\ref{fig1}), but instead represent the paths of  shielding current filaments. 

The green curve in Fig.~\ref{fig2} shows the safety-factor profile   calculated outside the magnetic separatrix in the presence of
the divertor plate. Note that the safety-factor is decreased in the presence of the  plate, because shielding 
current filaments can take a short-cuts through, rather than having to run in loops below, the plate. 

\begin{figure}[t]
\centerline{\includegraphics[width=\textwidth]{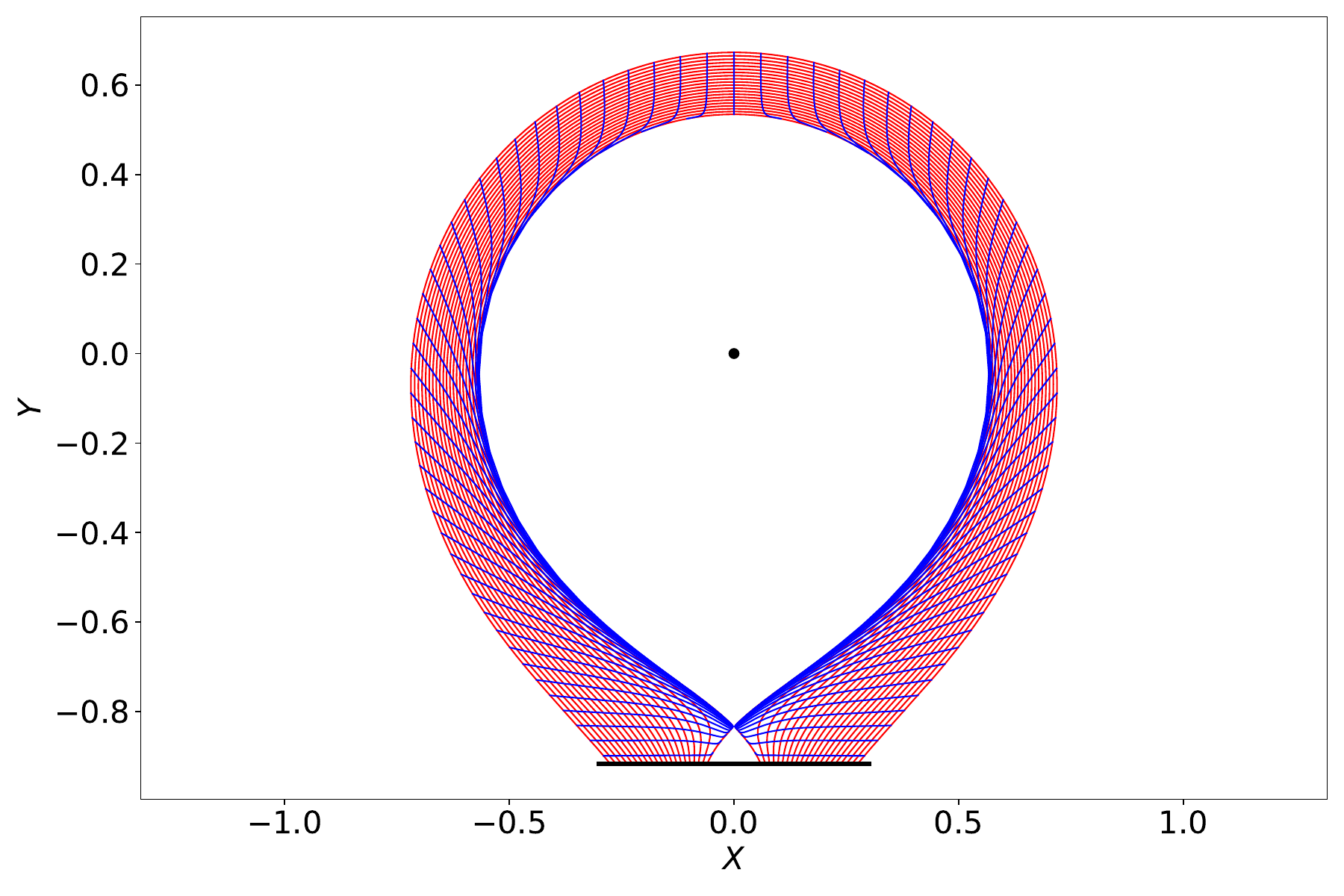}}
\caption{The straight-field-line coordinate system outside the magnetic separatrix in the presence of a divertor plate located halfway
between the X-point and the divertor coil filament. The red curves
are surfaces of constant $\hat{\psi}$, whereas the blue curves are surfaces of constant $\theta$. The black dot shows the location of the plasma current filament,
and the horizontal black line shows the location of the divertor plate. Here, $q_\ast=12$ and $\zeta=0.2$. }\label{fig6}
\end{figure}

Finally, Fig.~\ref{fig6}  shows the straight-field-line coordinate system  outside the magnetic separatrix in the presence of the divertor plate. Observe that the
divertor plate is a surface of constant $\theta$. 

\section{Improved model of a magnetically diverted plasma}\label{sthree}

\subsection{Introduction}
The model described in Sect.~\ref{stwo} has the unrealistic feature that $q=0$ at the magnetic axis. This feature is
an artifact of treating the plasma current as a filament, rather than a spatially extended distribution. In this
section, we shall attempt to construct an improved model.

\subsection{Equilibrium magnetic field}\label{equil}
Let $R$, $\phi$, $Z$ be conventional cylindrical coordinates that are coaxial with the plasma torus. Thus, $\nabla R\times\nabla\phi\cdot\nabla Z= 1/R$. 
We can  express the equilibrium magnetic field in the divergence-free manner \cite{rf2}
\begin{equation}
 {\bf B} = \nabla\phi\times \nabla\psi_p + B_0\,R_0\,g\,\nabla\phi= \nabla(\phi-q\,\theta)\times \nabla \psi_p.
 \end{equation}
Here, $\phi$ is a
true toroidal angle, $\psi_p$  the true poloidal magnetic flux (divided by $2\pi$), $g(\psi_p)$ an arbitrary (dimensionless) flux-function, $q(\psi_p)$ the safety-factor, $R_0$ the major radius of the
magnetic axis, and $B_0$  the vacuum toroidal magnetic field-strength on the axis (which implies that $g=1$ in the vacuum region surrounding the plasma). The previous equation is only self-consistent if 
\begin{equation}
\nabla\psi_p\times\nabla\theta\cdot\nabla\phi = \frac{B_0\,R_0\,g}{R^{\,2}\,q}.
\end{equation}

It is helpful to define a magnetic flux-surface label, $r$,  with the dimensions of length, as follows \cite{rf2}:
\begin{equation}\label{e30}
\frac{d\psi_p}{dr} =\frac{B_0\,r\,g}{q}.
\end{equation}
It is easily demonstrated that 
\begin{equation}\label{fluxc}
\nabla r\times\nabla\theta\cdot\nabla\phi = \frac{R_0}{R^{\,2}\,r}.
\end{equation}
Here, $r$ can be interpreted as the mean minor radius of a given magnetic flux-surface. Suppose that the magnetic
separatrix corresponds to the flux-surface $r=a$. Thus, we can interpret $a$ as the mean minor radius of the plasma. Moreover,
flux-surfaces with $r<a$ lie inside the separatrix, whereas those with $r>a$ lie outside. 

\subsection{Model safety-factor profile}
Let us adopt a model safety-factor profile inspired by the analysis of Sect.~\ref{stwo}. 
Let $\hat{r}=r/a$. Suppose that
\begin{align}\label{e37}
q(\hat{r}) = q_0 - \alpha_-\,\ln\left(1-\hat{r}^{\,2}\right)
\end{align}
for $\hat{r}<1$, and 
\begin{align}\label{e38}
q(\hat{r}) =- \alpha_+\,\ln\left(\hat{r}^{\,2}-1\right)
\end{align}
for $\hat{r}>1$, where
\begin{align}\label{e34}
\alpha_-& = -\frac{q_{95}-q_0}{\ln(1-\hat{r}_{95}^{\,2})},\\[0.5ex]
\alpha_+& = -\frac{q_{105}}{\ln(\hat{r}_{105}^{\,2}-1)}.\label{e35}
\end{align}
Here, $q_0$ is the safety-factor on the magnetic axis ($\hat{r}=0$), $q_{95}$ is the safety-factor on the magnetic flux-surface that
encloses $95\%$ of the poloidal magnetic flux enclosed by the magnetic separatrix, $\hat{r}_{95}$ is the $\hat{r}$-coordinate of the $95\%$
flux-surface, $q_{105}$ is the safety-factor on the magnetic flux-surface that
encloses $105\%$ of the poloidal magnetic flux enclosed by the magnetic separatrix, and $\hat{r}_{105}$ is the $\hat{r}$-coordinate of the $105\%$
flux-surface.

\subsection{Poloidal magnetic flux}
We wish to determine the poloidal flux in the vicinity of the magnetic separatrix. Let $\psi_p = B_0\,a^2\,\psi$. 
It follows  from Eq.~(\ref{e30})  that
\begin{equation}\label{e36}
\frac{d\psi}{d\hat{r}} \simeq \frac{\hat{r}}{q},
\end{equation}
assuming that we can set $g\simeq 1$ close to the separatrix. 

For $\hat{r}<1$, Eqs.~(\ref{e37}) and (\ref{e36}) yield
\begin{equation}
\frac{d\psi}{d\hat{r}} =\frac{\hat{r}}{q_0-\alpha_-\,\ln(1-\hat{r}^{\,2})},
\end{equation}
which can be integrated to give 
\begin{equation}
\psi(\hat{r}) = \frac{{\rm e}^{\,q_0/\alpha_-}}{2\,\alpha_-}\left\{E_1\!\left(\frac{q_0}{\alpha_-}\right)-E_1\!\left[\frac{q_0}{\alpha_-}-\ln\left(1-\hat{r}^{\,2}\right)\right]\right\},
\end{equation}
where $E_1(x)$ is an exponential integral \cite{as}. 
Hence,
\begin{equation}\label{psi1}
\psi(1) =  \frac{{\rm e}^{\,q_0/\alpha_-}}{2\,\alpha_-}\,E_1\!\left(\frac{q_0}{\alpha_-}\right)
\end{equation}
is the  poloidal magnetic flux (divided by $2\pi\,B_0\,a^2$) enclosed within the magnetic separatrix. 
Let us define the {\em normalized poloidal magnetic flux}, ${\mit\Psi}(\hat{r})=\psi(\hat{r})/\psi(1)$. Thus, ${\mit\Psi}=0$ on the magnetic axis, ${\mit\Psi}=1$ on the magnetic separatrix, and ${\mit\Psi}>1$ outside the separatrix. 
 It follows that
\begin{equation}
{\mit\Psi}(\hat{r})= 1- \frac{E_1[q_0/\alpha_--\ln(1-\hat{r}^{\,2})]}{E_1(q_0/\alpha_-)} = 1- \frac{E_1(q/\alpha_-)}{E_1(q_0/\alpha_-)}.
\end{equation}
By definition, ${\mit\Psi}(\hat{r}_{95}) = 0.95$, so 
\begin{equation}\label{e41}
 \frac{E_1(q_{95}/\alpha_-)}{E_1(q_0/\alpha_-)}=0.05.
\end{equation}
Assuming that $q_0$ and $q_{95}$ are specified, the  previous equation can be solved to give $\alpha_-$, which then allows  $\hat{r}_{95}$
to be determined from Eq.~(\ref{e34}).

For $\hat{r}>1$, Eqs.~(\ref{e38}) and (\ref{e36}) yield
\begin{equation}
\frac{d\psi}{d\hat{r}} = - \frac{\hat{r}}{\alpha_+\ln(\hat{r}^{\,2}-1)},
\end{equation}
which can be integrated to give
\begin{equation}
\psi(\hat{r}) = \psi(1) +\frac{E_1[-\ln(\hat{r}^{\,2}-1)]}{{2\,\alpha_+}}.
\end{equation}
Thus,
\begin{equation}\label{e44}
{\mit\Psi}(\hat{r}) = 1 + \frac{\alpha_-}{\alpha_+}\,{\rm e}^{-q_0/\alpha_-}\,\frac{E_1[-\ln(\hat{r}^{\,2}-1)]}{E_1(q_0/\alpha_-)}= 1+
 \frac{\alpha_-}{\alpha_+}\,{\rm e}^{-q_0/\alpha_-}\,\frac{E_1(q/\alpha_+)}{E_1(q_0/\alpha_-)},
\end{equation}
where use has been made of Eq.~(\ref{psi1}).
By definition, ${\mit\Psi}(\hat{r}_{105})=1.05$. Hence, Eqs.~(\ref{e41}) and (\ref{e44}) can be combined to give
\begin{equation}\label{e45}
 \frac{E_1(q_{105}/\alpha_+)}{\alpha_+}={\rm e}^{\,q_0/\alpha_-}\, \frac{E_1(q_{95}/\alpha_-)}{\alpha_-}.
\end{equation}
Assuming that $q_{105}$ is specified, the  previous equation can be solved to give $\alpha_+$, which then allows $\hat{r}_{105}$ to be determined
from Eq.~(\ref{e35}).

\begin{figure}[t]
\centerline{\includegraphics[width=0.8\textwidth]{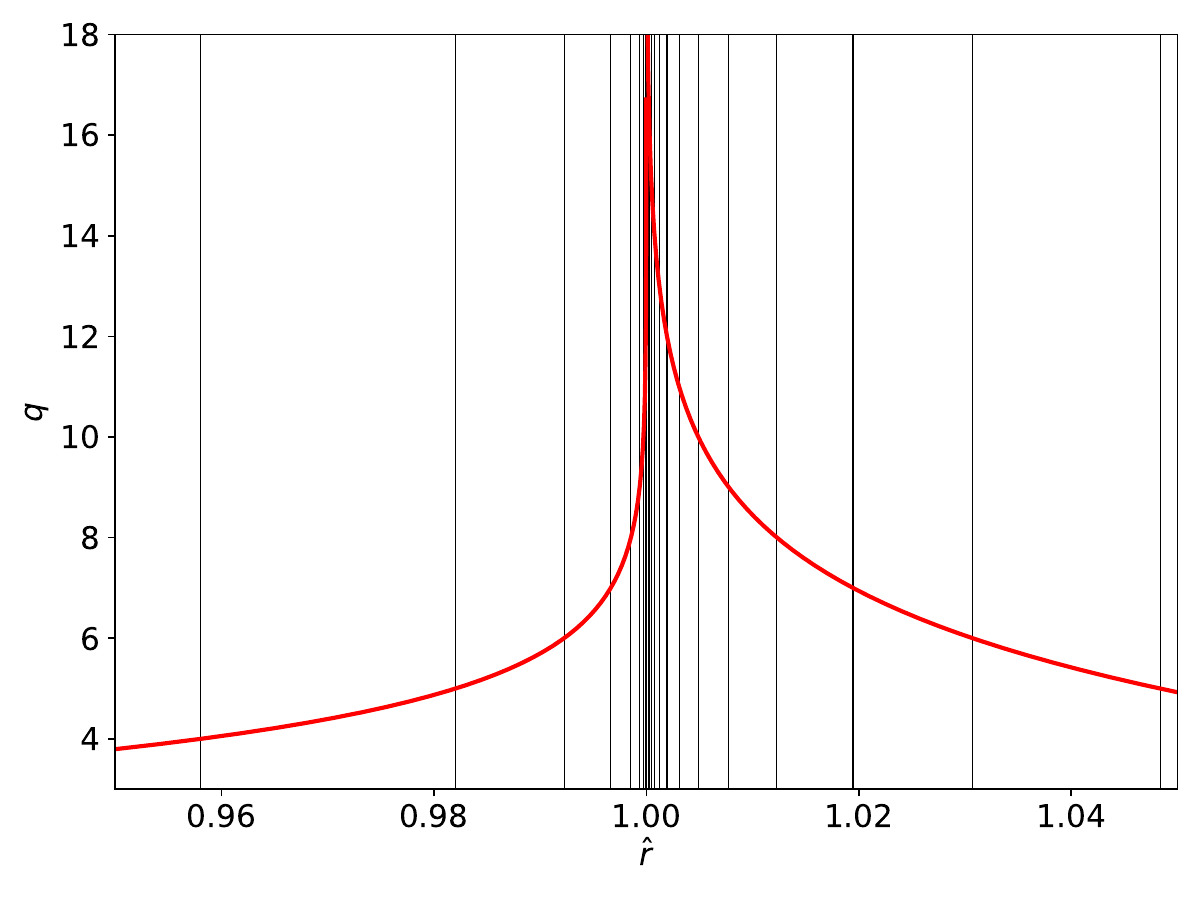}}
\caption{The model safety-factor profile in the vicinity of the magnetic separatrix, calculated for $q_0=1.01$,
$q_{95}=3.5$, and $q_{105}= 4.0$. The black vertical lines show the locations of the $n=1$ rational surfaces. }\label{fig7}
\end{figure}

Consider a plasma equilibrium characterized by $q_0= 1.01$, $q_{95}=3.5$, and $q_{105}=4.0$. We can solve Eqs.~(\ref{e41})
and (\ref{e45}) to give $\alpha_-= 1.196$ and $\alpha_+ = 2.163$. Equations~(\ref{e34}) and (\ref{e35}) then
reveal that $\hat{r}_{95} =0.9355$ and $\hat{r}_{105} =1.076$. The resulting model safety-factor profile in the vicinity of the
magnetic separatrix is shown in Fig.~\ref{fig7}. 

\section{Plasma response theory}\label{response}
\subsection{Introduction}
The aim of this section is analyze  the response of a tokamak plasma possessing a magnetic separatrix to a static RMP of toroidal mode number $n$. 

\subsection{Ideal-MHD perturbation}
Suppose that the plasma equilibrium is subject to a static magnetic perturbation that possesses $n$ periods in the toroidal direction, and is such that \cite{rf3}
\begin{equation}
\frac{r\,b^r(r,\theta,\phi)}{B_0\,R_0} = {\rm i}\left(\frac{R_0}{R}\right)^2
\sum_m\psi_m(r)\,{\rm e}^{\,{\rm i}\,(m\,\theta-n\,\phi)},
\end{equation}
where $b^r = {\bf b}\cdot\nabla r$,  ${\bf b}$ is the perturbed magnetic field, and the $\psi_m(r)$ are dimensionless functions. Note that the perturbation consists of a single toroidal
harmonic, but multiple  poloidal harmonics, characterized by the set of integer poloidal mode numbers $m$. As is well known, this is the case because, while
different toroidal harmonics are not coupled together in an axisymmetric plasma equilibrium, different poloidal
harmonics are coupled by the Shafranov shift and shaping of the equilibrium magnetic flux-surfaces \cite{rf2}. 
Everywhere in the plasma, apart from the immediate vicinity of the various rational surfaces, the perturbation is governed by the 
linearized equations of marginally-stable, ideal-magnetohydrodynamics (MHD) \cite{fkr}. These equations reduce to
a set of coupled ordinary differential equations (o.d.e.s) that
take the form \cite{rf2,rf3}
\begin{align}\label{e47}
r\,\frac{d\psi_m}{dr}&= \sum_{m'} \frac{L_m^{\,m'} \,Z_{m'}+M_{m}^{\,m'}\,\psi_{m'}}{m'-n\,q},\\[0.5ex]
(m-n\,q)\,r\,\frac{d}{dr}\!\left(\frac{Z_m}{m-n\,q}\right)&= \sum_{m'}\frac{N_m^{\,m'}\,Z_{m'} + P_m^{\,m'}\,\psi_{m'}}{m'-n\,q}.\label{e48}
\end{align}
Here, the (dimensionless) $Z_m(r)$ functions are related to the perturbed toroidal magnetic field. Moreover, the (dimensionless) $L_m^{\,m'}(r)$, $M_m^{\,m'}(r)$, $N_m^{\,m'}(r)$,
and $P_m^{\,m'}(r)$ coefficients are determined by the equilibrium profiles and the metric elements of the $r$, $\theta$, $\phi$ straight-field-line coordinate
system \cite{rf3}. 
Note that Eqs.~(\ref{e47}) and (\ref{e48}) are singular at rational magnetic flux-surfaces at which $q=m/n$.

The locations of the $n=1$ rational surfaces for our model safety-factor profile are  indicated in Fig.~\ref{fig7}. 
It can be seen that rational surfaces accumulate in the vicinity of the magnetic separatrix, where $q(\hat{r})$ has a logarithmic singularity. 
Moreover, rational surfaces exist both inside and outside the separatrix. 

\subsection{Behavior in vicinity of rational surface}
Consider the solution of the ideal-MHD o.d.e.s, (\ref{e47}) and (\ref{e48}), in the vicinity of a rational surface, of radius $r_s$, 
where $q(r_s)= m/n$, and $m$ is the resonant poloidal mode number. Let $x=r-r_s$. At small values of $|x|$, the function  $\psi_m(r)$ takes the
form \cite{rf2,rf3}
\begin{align}
\psi_m(r_s+x) = A_L^{\,\pm}\,|x|^{\,\nu_L}(1+\lambda_L\,x)+ A_S^{\,\pm}\,{\rm sgn}(x)\,|x|^{\,\nu_S}+ A_C\,x + {\cal O}(x^2),
\end{align}
where
\begin{align}\label{nul}
\nu_L &= \frac{1}{2}-\sqrt{-D_I},\\[0.5ex]
\nu_S&= \frac{1}{2}+\sqrt{-D_I},\label{nus}\\[0.5ex]
D_I &= - L_0\,P_0-\frac{1}{4},\\[0.5ex]
L_0 &= -\left(\frac{L_m^{\,m}}{m\,s}\right)_{r_s},\\[0.5ex]
P_0 &= -\left(\frac{P_m^{\,m}}{m\,s}\right)_{r_s},\\[0.5ex]
s &= \frac{d\ln q}{d\ln r},
\end{align}
and $\lambda_L$ and $A_C$ are defined in Refs.~\cite{rf2} and \cite{rf3}. The superscripts $+$ and $-$ correspond to $x>0$ and $x<0$, respectively. 
Here, $A_L^{\,\pm}$ are known as the {\em coefficients of the large solution}, whereas $A_S^{\,\pm}$ are termed  the {\em coefficients of the
small solution}. The {\em Mercier interchange stability parameter}, $D_I$, is assumed to be negative. (Otherwise, the plasma in the vicinity of the rational
surface would be unstable to localized ideal interchange modes \cite{mercier}.) Note that $\nu_L$ and $\nu_S$ are known as {\em Mercier indices}. 

In general, the response of the plasma in the vicinity of a given rational surface to the RMP is predominately {\em tearing parity}\/ in nature \cite{rf4}. 
For the case of a tearing parity response, the coefficients of the large solution to the left and the right of a rational surface are equal
to one another. In other words, $A_L^{\,+}=A_L^{\,-} = A_L$.  Note, however, that the coefficients of the small solution to the
left and the right of a rational surface are not, in general, equal to one another. 

\subsection{Plasma response equation}
Let us index the various rational surfaces in the plasma by means of an integer $k$. Note that this is possible even if there are an infinite number of
such surfaces. Thus, the $k$th rational surface possesses the minor radius $r_k$, the resonant poloidal mode number $m_k$, the
coefficient of the large solution $A_{L\,k}$, the coefficients of the small solution $A_{S\,k}^{\,\pm}$, and the Mercier indices 
$\nu_{L\,k}$ and $\nu_{S\,k}$. It is helpful to define the  quantities \cite{rf2,rf3}
\begin{align}
{\mit\Psi}_k &= r_k^{\,\nu_{L\,k}}\left(\frac{\nu_{S\,k} -\nu_{L\,k}}{L_{m_k}^{\,m_k}}\right)_{r_k}^{1/2}A_{L\,k},\\[0.5ex]
{\mit\Delta\Psi}_k &= r_k^{\,\nu_{S\,k}}\left(\frac{\nu_{S\,k} -\nu_{L\,k}}{L_{m_k}^{\,m_k}}\right)_{r_k}^{1/2}(A_{S\,k}^{\,+}-A_{S\,k}^{\,-}).
\end{align}
at each rational surface. Here, the complex dimensionless parameter ${\mit\Psi}_k$ is a measure of the driven reconnected magnetic flux
at the $k$th rational surface, whereas the complex dimensionless parameter ${\mit\Delta\Psi}_k$ is a measure of the strength of
the radially  localized shielding current that flows around the surface. 

The  response of a tokamak plasma to an RMP takes the general form \cite{rf2,rf3}
\begin{equation}\label{e58}
{\mit\Delta\Psi}_k= \sum_{k'}E_{kk'}\,{\mit\Psi}_{k'}+\chi_k,
\end{equation}
and specifies the shielding current that is excited in the vicinity of the $k$th rational surface in response to reconnected magnetic
flux at the other rational surfaces, as well as to the RMP. Here, $E_{kk'}$ is an Hermitian matrix that can be determined entirely from the ideal-MHD o.d.e.s, (\ref{e47})
and (\ref{e48}), subject to physical boundary conditions at small and large $r$ \cite{rf2,rf3}. Moreover, the real quantity $E_{kk}$
can be interpreted as the {\em tearing stability index}\/ \cite{fkr} of  a tearing mode that only reconnects magnetic flux at the
$k$th rational surface. Finally, the $\chi_k$ specify the poloidal harmonics of the RMP  \cite{rf3,rf4}. 

\subsection{Resistive layers}
In reality, the shielding current that flows in the vicinity of the $k$th rational surface does so in a resistive layer
whose thickness (in $r$), $\delta_k$, is much less than $r_k$ \cite{fkr}. We can define the complex dimensionless {\em layer response index}, 
\begin{equation}\label{e59}
{\mit\Delta}_k \equiv \frac{{\mit\Delta\Psi}_k}{{\mit\Psi}_k}, 
\end{equation}
which characterizes the tearing parity response of the plasma in the layer to the magnetic perturbation external to the layer. 
Equations~(\ref{e58}) and (\ref{e59}) can
be combined to give
\begin{equation}\label{e60}
\sum_{k'} ({\mit\Delta}_k\,\delta_{kk'} -E_{kk'})\,{\mit\Psi}_k= \chi_k,
\end{equation}
where $\delta_{kk'}$ is a Kronecker delta symbol. The previous equation is termed the {\em plasma response equation}, and specifies the 
reconnected magnetic
flux driven by the RMP at each rational surface in the plasma. 

Note that the derivation of Eq.~(\ref{e60}) depends crucially  on the principle of {\em asymptotic matching}\/ \cite{fkr}. According to this principle, 
the plasma response is governed by the equations of  linearized, marginally-stable, ideal-MHD throughout most of the plasma. However, these equations become
singular at the various rational surfaces in the plasma. [See Eqs.~(\ref{e47}) and (\ref{e48}).] The singularities are resolved by asymptotically matching
the ideal-MHD solution to resistive layer solutions at each rational surface in the plasma. 

\subsection{Ideal and vacuum responses}
Suppose, for the sake of argument, that the plasma only contains a single rational surface. In this case, the
plasma response equation, (\ref{e60}),  reduces to
\begin{equation}
{\mit\Psi}_k = \frac{\chi_k}{{\mit\Delta}_k - E_{kk}}.
\end{equation}
Alternatively, Eq.~(\ref{e58}) yields
\begin{equation}
{\mit\Delta\Psi}_k = E_{kk}\,{\mit\Psi}_k + \chi_k.
\end{equation}
Let us  suppose that $E_{kk}<0$: i.e., the tearing mode resonant at the surface is intrinsically stable. In this case, any magnetic
reconnection that takes place at the surface is entirely due to the RMP.
We can distinguish two asymptotic limits. In the first limit, $|{\mit\Delta}_k|\gg (-E_{kk})$. In this limit, the shielding current excited at the
rational surface is sufficiently strong that driven magnetic reconnection is largely suppressed at the surface \cite{rf1,rf2}. In other words,
${\mit\Psi}_k\simeq 0$. We term this an {\em ideal response}, because it is the exactly the same response as one would get by imposing the
ideal-MHD flux-freezing  constraint that the topology of the magnetic field cannot change at the rational surface: i.e., ${\mit\Psi}_k=0$ \cite{boozer}. 
In the second limit, $|{\mit\Delta}_k|\ll (-E_{kk})$. In this limit, the shielding current is too weak to prevent
driven magnetic reconnection from occurring at the rational surface. Thus, ${\mit\Psi}_k\simeq \chi_k/(-E_{kk})$. 
We term this a {\em vacuum response}, because it is exactly the same response as one would get by imposing the
vacuum constraint that no shielding current flows at the rational surface: i.e., ${\mit\Delta\Psi}_k=0$. 

In a region of the plasma in which the response to the RMP lies in the ideal regime [i.e., $|{\mit\Delta}_k|/(-E_{kk})\gg 1$ at all rational surfaces], the perturbed magnetic field has
a strong dependence on the equilibrium magnetic field, because the equilibrium field determines the locations of the rational
surfaces on which the constraints ${\mit\Psi}_k= 0$ have to be imposed. On the other hand, in a region of the plasma in which the response lies in the vacuum
regime [i.e., $|{\mit\Delta}_k|/(-E_{kk})\ll 1$ at all rational surfaces], the perturbed magnetic field has no (local) dependence on the equilibrium magnetic field. This is the case because we
can write the perturbed field in a current-free region  in the form ${\bf b} = \nabla V$, where the vacuum potential satisfies
$\nabla^2 V=0$, which completely independent of the equilibrium field. Moreover, writing the perturbed magnetic field in this form
automatically satisfies the constraints that ${\mit\Delta\Psi}_k=0$ at the rational surfaces. Thus, in the vacuum regime, there is no need for us to determine the locations of the
rational surfaces, which means that we do not have to employ a straight-field-line coordinate system, which implies that there is no reason for our
adopted coordinate system to become singular anywhere. 

\section{Behavior close to magnetic separatrix}\label{separatrix}
\subsection{Introduction}
The aim of this section is to investigate asymptotic matching in the immediate vicinity of the magnetic separatrix. 

\subsection{Spacing between successive rational surfaces}
Suppose that $\hat{r}_{k+1}>\hat{r}_k$ for all values of $k$, where $\hat{r}_k=r_k/a$.  In other words, suppose that the rational surfaces are indexed in order of increasing minor radius.
In the vicinity of the magnetic separatrix, Eqs.~(\ref{e37}) and (\ref{e38}) yield
\begin{equation}
\hat{r}_k\simeq \left[1-\exp\left(-\frac{m_k}{n\,\alpha_-}\right)\right]^{1/2}
\end{equation}
for $\hat{r}_k<1$, and 
\begin{equation}
\hat{r}_k= \left[1+\exp\left(-\frac{m_k}{n\,\alpha_+}\right)\right]^{1/2}
\end{equation}
for $\hat{r}_k>1$. 

The spacing (in $\hat{r}$) between successive rational surfaces is
\begin{equation}
\hat{\epsilon}_k(\hat{r}_k) = \frac{d\hat{r}_k}{dm_k}
\end{equation}
for $\hat{r}_k<1$, 
and
\begin{equation}
\hat{\epsilon}_k(\hat{r}_k) =- \frac{d\hat{r}_k}{dm_k}
\end{equation}
for $\hat{r}_k>1$.
Thus,
\begin{equation}
\hat{\epsilon}_k(\hat{r}_k)= \frac{1-\hat{r}_k^{\,2}}{2\,n\,\alpha_-\,\hat{r}_k}
\end{equation}
for $\hat{r}_k<1$, and 
\begin{equation}
\hat{\epsilon}_k(\hat{r}_k)= \frac{\hat{r}_k^{\,2}-1}{2\,n\,\alpha_+\,\hat{r}_k}
\end{equation}
for $\hat{r}_k>1$. 
Note that the spacing between successive rational surfaces tends to zero linearly as the magnetic separatrix ($\hat{r}_k=1$) is approached. 

\subsection{Magnetic shear}\label{shear}
In the vicinity of the magnetic separatrix, the magnetic shear can be written
\begin{equation}\label{e69}
s(\hat{r}) \equiv \frac{\hat{r}}{q}\,\frac{dq}{d\hat{r}}\simeq -\frac{1}{(\hat{r}-1)\,\ln(2\,|\hat{r}-1|)}.
\end{equation}
Observe that the magnitude of the shear becomes infinite as the magnetic separatrix ($\hat{r}=1$) is approached. 

\subsection{Resistive layer quantities}
It is helpful to define the following quantities that parameterize a general resistive layer \cite{layer}:
\begin{align}
\ln{\mit\Lambda} &= 24 - \ln\left[\left(\frac{n_e}{10^6}\right)^{1/2}\left(\frac{e}{T_e}\right)\right],\\[0.5ex]
\tau_{ei} &= \frac{6\sqrt{2}\,\pi^{3/2}\,\epsilon_0^{\,2}\,m_e^{1/2}\,T_e^{\,3/2}}{Z\,\ln{\mit\Lambda}\,e^4\,n_e},\\[0.5ex]
\eta_\parallel &= \frac{m_e}{1.96\,n_e\,e^2\,\tau_{ei}},\\[0.5ex]
\tau_R &= \frac{\mu_0\,r^2}{\eta_\parallel},\\[0.5ex]
\tau_A &= \frac{R_0\,\sqrt{\mu_0\,m_i\,n_e}}{B_0},\\[0.5ex]
\tau_E &= \frac{r^2}{\chi_E},\\[0.5ex]
\tau_\phi &= \frac{r^2}{\chi_\phi},\\[0.5ex]
d_\beta &= \frac{\sqrt{(5/3)\,m_i\,(T_e+T_i)}}{e\,B_0},\\[0.5ex]
S &= \frac{\tau_R}{\tau_A}.
\end{align}
Here, $Z$ is the effective ion charge number, $T_e(r)$  the electron temperature, $T_i(r)$  the ion temperature, $m_e$ the electron mass,  
$m_i$ the ion mass, $\chi_E(r)$ the perpendicular energy
diffusivity, and $\chi_\phi(r)$ the perpendicular momentum diffusivity. Furthermore, $\tau_R(r)$ is the resistive diffusion time,
$\tau_A(r)$ the Alfv\'{e}n time, $\tau_E(r)$ the energy confinement time,  $\tau_\phi(r)$ the momentum confinement time,  $d_\beta(r)$
the ion sound-radius, and $S(r)$ the Lundquist number. 

\subsection{Resistive layer equation}\label{layer}
Let $Y(r)\,\exp[\,{\rm i}\,(m\,\theta-n\,\phi)]$ represent the perturbed electron fluid stream-function in the laboratory frame. Consider the resistive layer centered on the $k$th rational surface. Let
$X=S^{\,1/3}\,(r-r_k)/r_k$, where $S$ is evaluated at radius $r_k$. We can write
\begin{equation}
\overline{Y}(p)= \int_{-\infty}^\infty Y(X)\,{\rm e}^{-{\rm i}\,p\,X}\,dX.
\end{equation}
The boundary conditions are that $\overline{Y}(p)\rightarrow 0$ as $p\rightarrow\infty$, and 
\begin{equation}\label{e79}
\overline{Y}(p)=Y_0 \left[\frac{S^{-1/3}\,{\mit\Delta}_k}{\pi\,p} + 1+{\cal O}(p)\right]
\end{equation}
at small values of $p$ \cite{layer,cole}. Here, $Y_0$ is an arbitrary constant, and ${\mit\Delta}_k$ is the complex (dimensionless) layer response index introduced in Eq.~(\ref{e59}). 

Suppose that the resistive layer physics is controlled by the low-$\beta$,  three-field, extended-MHD model described in Ref.~\cite{layer}. In this
case, the Fourier transformed layer equations reduce to the following equation:
\begin{equation}\label{e80}
\frac{d}{dp}\!\left[A(p)\,\frac{d\overline{Y}}{dp}\right] - \frac{B(p)}{C(p)}\,p^2\,\overline{Y}=0,
\end{equation}
where
\begin{align}
A(p) &= \frac{(n\,s)^2\,p^2}{{\rm i}\,(Q_E+Q_e)+p^2},\\[0.5ex]
B(p)&= - Q_E\,(Q_E+Q_i)+{\rm i}\,(Q_E+Q_i)\,(P_\phi+P_E)\,p^2 + P_\phi\,P_E\,p^4,\\[0.5ex]
C(p) &= {\rm i}\,(Q_E+Q_e)+ \left[P_E+{\rm i}\,(Q_E+Q_i)\,D^{\,2}\right]p^2
+ (1+1/\tau)\,P_\phi\,D^{\,2}\,p^4.\label{e83}
\end{align}
Here,
\begin{align}
\tau &=-\frac{ \omega_{\ast\,e}}{\omega_{\ast\,i}},\\[0.5ex]
Q_E &= -S^{1/3}\,n\,\omega_E\,\tau_A,\\[0.5ex]
Q_{e,i} &=- S^{1/3}\,n\,\omega_{\ast\,e,i}\,\tau_A,\\[0.5ex]
D &= S^{1/3}\left(\frac{\tau}{1+\tau}\right)^{1/2}\hat{d}_\beta,\\[0.5ex]
P_E&= \frac{\tau_R}{\tau_E},\\[0.5ex]
P_\phi&= \frac{\tau_R}{\tau_\phi},
\end{align}
and
\begin{equation}
\omega_{\ast\,i}(r) = -\frac{1}{e\,n_e}\,\frac{dp_i}{d\psi_p}
\end{equation}
is the {\em ion diamagnetic frequency}. Moreover, $p_i(r)$ is the ion pressure, and $\hat{d}_\beta=d_\beta/r$. As before, all quantities are evaluated at radius $r_k$. 

\subsection{Solution of resistive layer equation}\label{layer1}
As we saw in Sect.~\ref{shear}, the magnetic shear becomes very large as the magnetic separatrix is approached. If we treat $(n\,s)^2$ as much
larger than the other parameters (e.g., $Q_E$, $Q_e$, $P_E$, $D$) that appear in Eqs.~(\ref{e80})--(\ref{e83}) then it is clear that the first term on the left-hand side of Eq.~(\ref{e80}) dominates the second term in the region  $p\lesssim1$. In this case, we can integrate  the equation to give
\begin{equation}
\overline{Y}(p)= Y_0\left\{\frac{S^{-1/3}\,{\mit\Delta}_k}{\pi}\left[\frac{1}{p} + \frac{p}{-{\rm i}\,(Q_E+Q_e)}\right] + 1+{\cal O}(p^2)\right\}
\end{equation}
for $p\lesssim 1$, where use has been made of Eq.~(\ref{e79}). Incidentally, the previous equation bears the hallmark of a {\em constant-$\psi$} layer response
regime \cite{fkr,layer}. 

Let $p=(n\,|s|)^{1/2}\,\hat{p}$. For $\hat{p}\gtrsim 1$, Eqs.~(\ref{e80})--(\ref{e83}) yield
\begin{equation}
\frac{d^2\overline{Y}}{d\hat{p}^{\,2}} -\frac{B(\hat{p})}{C(\hat{p})}\,\hat{p}^{\,2}\,\overline{Y}\simeq 0,
\end{equation}
where 
\begin{align}
B(\hat{p}) &= - Q_E\,(Q_E+Q_i)+{\rm i}\,(Q_E+Q_i)\,(P_\phi+P_E)\,(n\,|s|)\,\hat{p}^{\,2} + P_\phi\,P_E\,(n\,|s|)^2\,\hat{p}^{\,4}\nonumber\\[0.5ex]
&\simeq  P_\phi\,P_E\,(n\,|s|)^2\,\hat{p}^{\,4},\\[0.5ex]
C(\hat{p}) &=  {\rm i}\,(Q_E+Q_e)+ \left[P_E+{\rm i}\,(Q_E+Q_i)\,D^{\,2}\right](n\,|s|)\,\hat{p}^{\,2}
+ (1+1/\tau)\,P_\phi\,D^{\,2}\,(n\,|s|)^2\,\hat{p}^{\,4}\nonumber\\[0.5ex]
&\simeq (1+1/\tau)\,P_\phi\,D^{\,2}\,(n\,|s|)^2\,\hat{p}^{\,4},
\end{align}
which reduces to 
\begin{equation}
\frac{d^2\overline{Y}}{dp^2} - G\,p^2\,\overline{Y}\simeq 0,
\end{equation}
where 
\begin{equation}
G = \frac{P_E}{(1+1/\tau)\,D^{\,2}\,n^2\,|s|^2}= \frac{\tau_A^{\,2/3}\,\tau_R^{\,1/3}}{\tau_E\,\hat{d}_\beta^{\,2}\,(n\,|s|)^2}.
\end{equation}

In the language of Ref.~\cite{layer}, we have found a constant-$\psi$ layer solution characterized by  $\nu=1/4$. 
It follows from the analysis of Ref.~\cite{layer} that
\begin{equation}
S^{-1/3}\,{\mit\Delta}_k = \frac{\nu^{\,2\nu-1}\pi\,\Gamma(1-\nu)}{\Gamma(\nu)}\,[\,{\rm i}\,(Q_E+Q_e)]\,G^{\,\nu},
\end{equation}
where $\Gamma(z)$ is a gamma function \cite{as}.
Hence, in the continuum limit in which the resistive layers are very closely spaced, 
\begin{equation}
{\mit\Delta}_k(\hat{r})= -{\rm i}\,\frac{2\pi\,\Gamma(3/4)}{\Gamma(1/4)} \,
\frac{n\,(\omega_E+\omega_{\ast\,e})\,\tau_A^{\,1/2}\,\tau_R^{\,3/4}}{\tau_E^{\,1/4}\,\hat{d}_\beta^{\,1/2}\,(n\,|s|)^{1/2}},
\end{equation}
where all terms on the right-hand side are evaluated at the normalized radius $\hat{r}$. Thus, we deduce that the strong magnetic shear in
the vicinity of the magnetic separatrix forces all of the resonant layers in this region to lie in the so-called {\em diffusive-resistive}\/ regime 
introduced in Ref.~\cite{layer}.

Finally, the thickness of  a given resistive layer in $p$-space is $|G|^{-\nu}$ \cite{layer}, so the thickness in $\hat{r}$-space is
\begin{equation}
\hat{\delta}_k(\hat{r}) = \frac{\tau_A^{\,1/2}}{\tau_R^{\,1/4}\,\tau_E^{\,1/4}\,\hat{d}_\beta^{\,1/2}\,(n\,|s|)^{1/2}}.
\end{equation}

\subsection{Mercier indices}
The resistive layer solutions discussed in Sects.~\ref{layer} and \ref{layer1} are premised on the assumption that the
Mercier indices at the rational surfaces, $\nu_{L\,k}$ and $\nu_{S\,k}$,  take the respective values $0$ and $1$. 
As is described in Ref.~\cite{rf5}, if this is not the case then the mismatch between the Mercier indices in the layer
and those in the outer region (i.e., the region of the plasma that is governed by ideal-MHD) can be reconciled by matching the layer to the
outer solution by means of an intermediate layer. However, the cylindrical expression for the Mercier interchange stability parameter takes the form 
\begin{equation}
D_{I\,k}=-\frac{\mu_0}{B_0^{\,2}}\left[\frac{2\,(1-q^2)}{s^2}\,\hat{r}\,\frac{dp}{d\hat{r}}\right]-\frac{1}{4},
\end{equation}
where $p(\hat{r})$ is the total plasma pressure, and all quantities are evaluated at the rational surface \cite{ggj}. 
Although the magnetic separatrix, $\hat{r}=1$,  is not well-described by the cylindrical approximation, if we use the previous expression to roughly gauge how
$D_{I\,k}$ varies in the vicinity of the separatrix then the fact that $q\sim \ln|\hat{r}-1|$, whereas
 $|s|\sim 1/(|\hat{r}-1|\,\ln|\hat{r}-1|)$ [see Eqs.~(\ref{e38}) and (\ref{e69})], indicates that the strong magnetic shear in the vicinity of the
 separaratrix ensures that $D_{I\,k}\rightarrow -1/4$.
 In this  case, $\nu_{L\,k}\rightarrow 0$ and $\nu_{S\,k}\rightarrow 1$  [see
 Eqs.~(\ref{nul}) and (\ref{nus})], and there is no need for an intermediate layer. 
 
\begin{figure}[t]
\centerline{\includegraphics[width=0.6\textwidth]{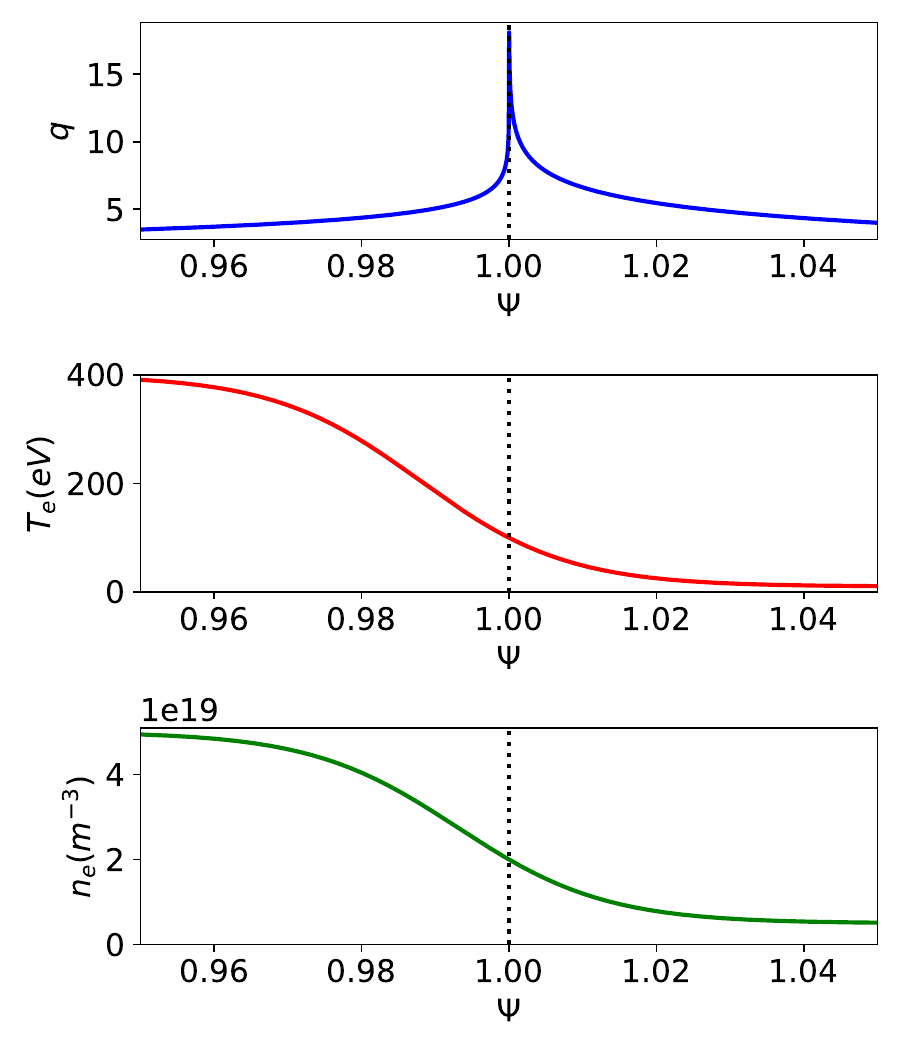}}
\caption{Model safety-factor, electron temperature, and electron number density profiles at the edge of a typical JET  H-mode plasma. 
The vertical dotted lines show the location of the magnetic separatrix. }\label{fig8}
\end{figure}

\subsection{Plasma parameters}
 Suppose that the safety-factor profile is characterized by $q_0=1.01$, $q_{95}=3.5$, and $q_{105}=4.0$, as  was previously assumed. 
 
 We shall adopt the model edge electron temperature and 
 number density profiles shown in Fig.~\ref{fig8}. These are modified hyperbolic tangent ({\rm mtanh}) profiles deduced from
 Figs.~1 and 2 of Ref.~\cite{jet}, and represent a typical reactor-relevant high-density/low-temperature pedestal in a   JET H-mode plasma (discharge 84600). We shall assume that
 $T_i=T_e$, for the sake of simplicity. 
 
\begin{figure}[t]
\centerline{\includegraphics[width=0.6\textwidth]{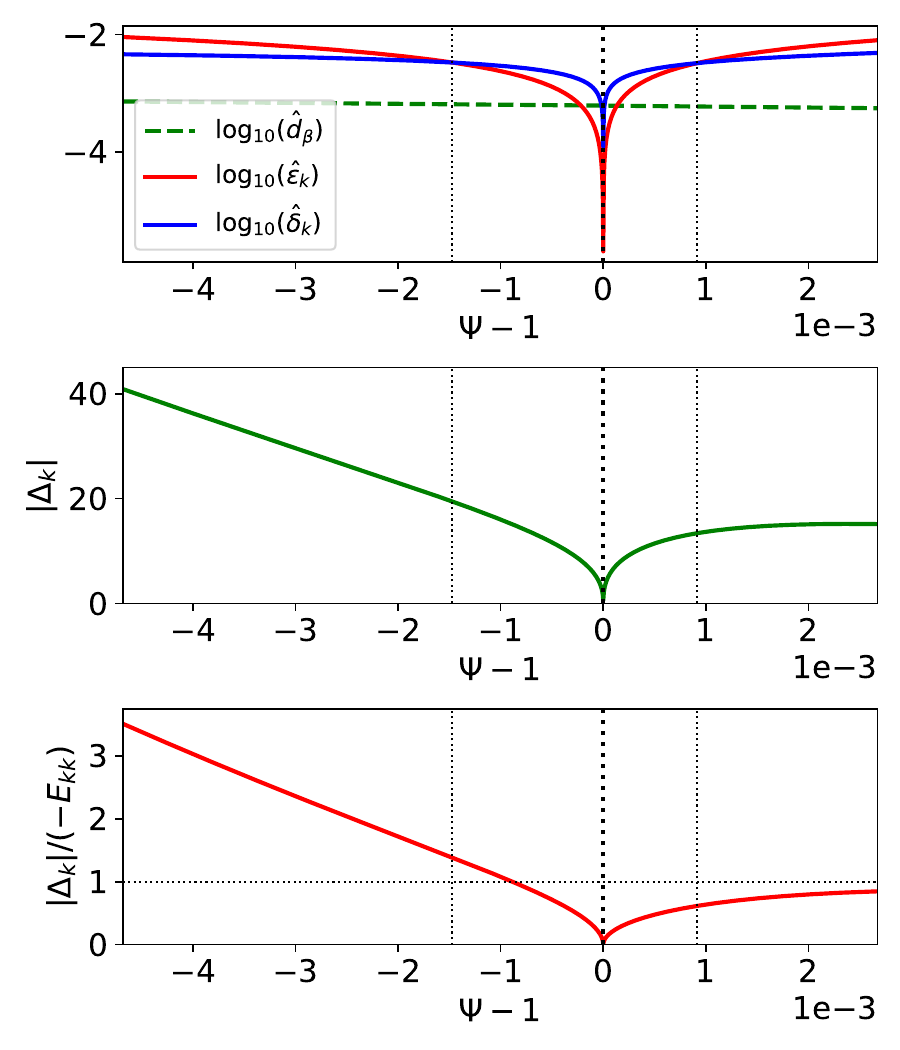}}
\caption{Overlap criterion for the resistive layers associated with an $n=1$ RMP close to the magnetic
separatrix of the JET plasma shown in Fig.~\ref{fig8}. The bold vertical dotted lines show the location of the magnetic separatrix, whereas
the faint vertical dotted lines indicate the boundaries of the region of resistive layer overlap.}\label{fig9}
\end{figure}

Let us suppose that $\omega_E+ \omega_{\,\ast\,e}\simeq \omega_{\ast\,e}$ close to the magnetic separatrix. This is a reasonable assumption because we have no reason to imagine that the magnitude of $\omega_E$ greatly exceeds that of $\omega_{\ast\,e}$
 in the pedestal region of an H-mode tokamak plasma. In fact, Fig.~3 of Ref.~\cite{diiid} suggests that the magnitude of 
 $\omega_E$ is about a factor of 5 times smaller than that of $\omega_{\ast\,e}$ in the pedestal of a typical DIII-D H-mode plasma (discharge
 158115). 
 
 We can complete our model by adopting the following JET-like values of the remaining unspecified plasma parameters: $B_0=3.45\,{\rm T}$, $R_0=2.96\,{\rm m}$, $a=1.25\,{\rm m}$, $Z=10.0$,
 $M=2.0$, $\chi_E=\chi_\phi=1\,{\rm m^{2}/s}$ \cite{jetx}. Here, $M$ is the ion mass number, and the rather large value of $Z$ is meant to account for the presence of a significant fraction
 of trapped particles, in addition to plasma impurities, close to the plasma boundary. 

 \begin{figure}[t]
\centerline{\includegraphics[width=0.6\textwidth]{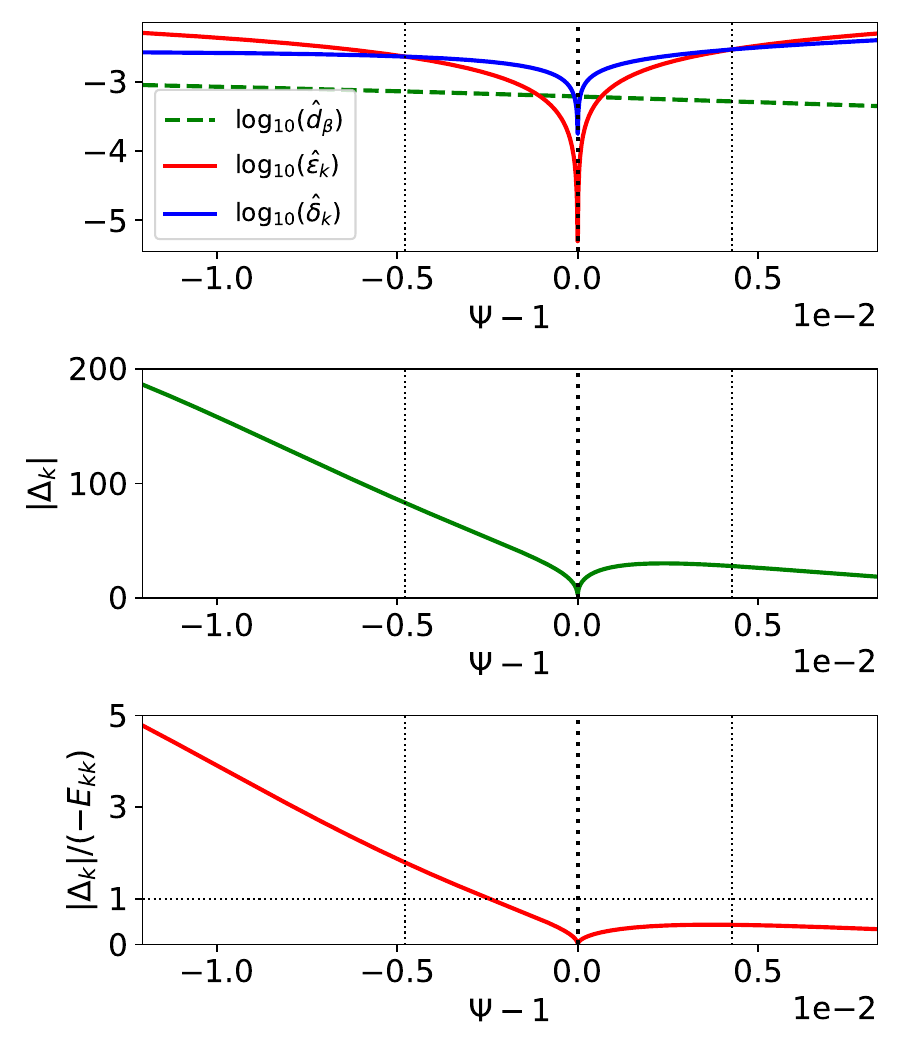}}
\caption{Overlap criterion for the resistive layers associated with an $n=4$ RMP close to the magnetic
separatrix of the JET plasma shown in Fig.~\ref{fig8}. The bold vertical dotted lines show the location of the magnetic separatrix, whereas
the faint vertical dotted lines indicate the boundaries of the region of resistive layer overlap.}\label{fig10}
\end{figure}
 
\subsection{Overlap of resistive layers}
The top panel of Fig.~\ref{fig9} shows the spacing (in $\hat{r}$) between successive rational surfaces $\hat{\epsilon}_k({\mit\Psi})$, as well as the resistive 
layer width  (in $\hat{r}$), $\hat{\delta}_k({\mit\Psi})$, calculated in the vicinity of the magnetic separatrix for the case of an $n=1$ RMP applied to our model JET equilibrium. 
It can be seen that the spacing between rational surfaces and the resistive layer width both tend to zero as the separatrix (${\mit\Psi}=1$) is approached. However, the rational surface spacing tends to zero faster than the layer width. Consequently, the resistive layers overlap one another
in the immediate vicinity of the separatrix. In fact, the region of overlap extends from ${\mit\Psi}=0.9985$ to $1.0009$. 
Note, from the figure, that the resistive layers overlap well before either the spacing between successive rational surfaces or the layer thickness
fall below the ion gyroradius (which is approximately equal to $d_\beta$). 

The middle panel of Fig.~\ref{fig9} shows the magnitude of the layer response index, $|{\mit\Delta}_k|$. It can be seen that this quantity is zero
at the separatrix, but increases steeply as we move from the separatrix into the interior of the plasma, indicating that strong
shielding currents are excited in the plasma interior. The steep increase is mostly due to the fact that the magnetic shear exhibits a
steep decrease as we move away from the separatrix.
On the other hand, the magnitude of the layer response index seems to
asymptote to a constant value as we move from the separatrix into the scape-off layer (SOL) (i.e., the region ${\mit\Psi}>1$), indicating that only comparatively weak
shielding currents are excited in the SOL. The magnitude of the index fails to exhibit a steep increase as we move into the SOL,
despite the fact that the magnetic shear is strongly decreasing, because the electron temperature falls off rapidly with increasing distance from the separatrix.
 In reality, the value of $|{\mit\Delta}_k|$ in the SOL is probably an overestimate, because we have not taken into account the fact that shielding current filaments in the SOL run partially through the divertor plate, which is likely to increase the effective resistivity of rational magnetic flux-surfaces in this region. 

The lower panel of Fig.~\ref{fig9} shows the ratio $|{\mit\Delta}_k|/(-E_{kk})$. Here, we have used the cylindrical result $E_{kk}=-2\,|m|$
to estimate $E_{kk}$. It is clear that $|{\mit\Delta}_k|/(-E_{kk})\gg 1$ in the interior of the plasma. In other words, the plasma interior 
exhibits an ideal response to the RMP. However, $|{\mit\Delta}_k|/(-E_{kk})\lesssim 1$ in the overlap region and the SOL. In other words,  
it would not be unreasonable to characterize the response of the plasma in the overlap region and the  SOL as being vacuum-like. 
 Note that it is not a coincidence that the inner boundary of the overlap region corresponds approximately
to the point where $|{\mit\Delta}_k|/(-E_{kk})=1$. The plasma at the separatrix is sufficiently hot, with a typical electron temperature of 100\,eV, that
$|{\mit\Delta}_k|/(-E_{kk})$ would greatly exceed unity were $|s|\sim {\cal O}(1)$. The same factor that causes the overlap of
resistive layers also causes $|{\mit\Delta}_k|/(-E_{kk})$ to dip below unity very close to the separatrix, and this factor is, of course,  the very strong magnetic
shear in the immediate vicinity of the separatrix.

Figure~\ref{fig10} shows analogous data to that displayed in Fig.~\ref{fig9} for the case of an $n=4$ RMP. In this case, the region of resistive layer overlap
extends from ${\mit\Psi}= 0.9952$ to $1.0043$. Thus, the width  in ${\mit\Psi}$ of the overlap region exhibits a roughly linearly scaling with $n$. 
As before, the response of the plasma interior is ideal, whereas that of the overlap region and the  SOL could reasonably
be characterized as being vacuum-like,

\subsection{Discussion}
We have seen that rational surfaces accumulate in the vicinity of the magnetic separatrix. Moreover, the resistive layers, 
centered on the various rational surfaces, merge into one another in a very narrow region that straddles the  separatrix. 
Furthermore, while the plasma response in the plasma interior lies in the ideal regime, that in the region of resistive layer
overlap and the SOL can reasonably be modeled as being vacuum-like. It follows that the magnetic perturbation in the plasma interior is governed by Eqs.~(\ref{e47}) and (\ref{e48}). In order to solve these equations, we need to know the locations of all of the rational magnetic flux-surfaces
in this region, which implies that we must adopt a straight-field-line coordinate system. On the other hand, in the overlap region and the SOL, we can
write ${\bf b}=\nabla V$, where $\nabla^2 V=0$, which automatically ensures that the response is vacuum-like. In this case, there is
no need to know the locations of the rational surfaces, which implies that it is not necessary to adopt a straight-field-line coordinate system. 

We are now in a position to formulate a practical, yet physically justified,  approach to calculating the response of a magnetically
diverted tokamak plasma to an RMP. Let the inner boundary of the region of resistive layer overlap that straddles the 
magnetic separatrix lie at ${\mit\Psi}=1-\epsilon_c$. Note that $0<\epsilon_c\ll 1$. We must solve Eqs.~(\ref{e47}) and (\ref{e48}) in the region $0<{\mit\Psi}<1-\epsilon_c$, using
the straight-field-line coordinate system specified in Eq.~(\ref{fluxc}), and treating the region ${\mit\Psi}>1-\epsilon_c$ as a vacuum. Of course, this
is the approach taken in existing plasma response codes such as GPEC \cite{ipec,park,gpec,logan}, in which $\epsilon_c$ is given a small positive value that is determined via an empirical guideline.  (This course of action is adopted because GPEC cannot deal with the singularity of the straight-field-line coordinate system at ${\mit\Psi}=1$, and
is taken in the hope that the neglect of a very narrow strip of plasma lying just inside the separatrix,
as well as the plasma in the SOL,  does not unduly affect the results of the calculation.) However, the analysis of this paper allows us to
determine the correct value of $\epsilon_c$ from the edge plasma parameters, and also justifies the neglect of the plasma in the region ${\mit\Psi}>1-\epsilon_c$.  The fact that the region ${\mit\Psi}>1-\epsilon_c$ is treated as a vacuum
means that rational surfaces in this region do not contribute to the plasma response. In other words, the plasma response
calculation only needs to take a finite, rather than an infinite,  number of rational surfaces into account. Indeed, for an $n=1$ RMP, applied to our model JET equilibrium, we
only need to include rational surfaces whose resonant poloidal mode numbers are less than 8. On the other hand, for an $n=4$ RMP, we  need to include all rational surfaces whose resonant poloidal mode numbers are less than 24. 
Because the straight-field-line coordinate
system is only required in the region $0<{\mit\Psi}<1-\epsilon_c$,  where $\epsilon_c>1$,  the fact that the system becomes singular at ${\mit\Psi}=1$ is irrelevant. Indeed, we are
free to choose a convenient non-singular coordinate system in the vacuum region, ${\mit\Psi}>1-\epsilon_c$. 

If we suppose that $q_{95}$ is gradually increased then, at some stage, a new rational surface will enter the domain of solution, $0<{\mit\Psi}<1-\epsilon_c$,
from the region of resistive layer overlap. Suppose that the new surface is the $k$th surface. The entry of the surface into the domain of solution
will not cause the sudden imposition of the additional ideal constraint ${\mit\Psi}_k=0$, which would also cause a sudden change in the plasma
response to the RMP \cite{zheng}, because $|{\mit\Delta}_k|/(-E_{kk})$ is initially quite small, due to the large magnetic shear at ${\mit\Psi}=1-\epsilon_c$. However, 
as $q_{95}$ is further increased, the new surface will move into the plasma interior, where the shear is moderate, and the ideal constraint ${\mit\Psi}_k=0$
will eventually hold. In other words, ideal constraint ${\mit\Psi}_k=0$ develops gradually, as the new surface moves into the plasma interior,
rather than being imposed immediately, as soon as the surface enters the domain of solution. 

\section{Summary}\label{sum}
In Sect.~\ref{stwo}, we construct a simple model of a magnetically diverted tokamak plasma. We argue that the commonly used term ``last closed magnetic flux-surface" (LCFS) is misleading. 
In fact, all equilibrium magnetic flux-surfaces are closed. However, flux-surfaces that lie outside the magnetic separatrix (i.e., the flux-surface that
contains the X-point) are only partially occupied by plasma. We show that it is possible to calculate  unique values of the safety-factor 
on flux-surfaces that lie both inside and outside the magnetic separatrix. We conclude  that rational magnetic flux-surfaces exist in both regions.  The safety-factor is shown to diverge logarithmically as the magnetic separatrix is approached. 

 A more realistic model of a magnetically diverted 
tokamak plasma is introduced in Sect.~\ref{sthree}. 

In Sect.~\ref{response}, we show that the response of a magnetically diverted tokamak plasma to an externally generated, static,  RMP can be formulated
as an asymptotic matching problem. The response throughout most of the plasma is governed by the equations of
linearized,  marginally-stable, ideal-MHD. However, these equations are singular at the various rational surfaces in the plasma. 
The singularities are resolved by asymptotically matching the ideal-MHD solution to resistive layer solutions centered on the rational
surfaces. The fact that the safety-factor diverges logarithmically as the magnetic separatrix is approached means that, in principle, the asymptotic matching problem involves an
infinite number of coupled rational surfaces, the majority of which are located very close to the magnetic separatrix. 

In Sect.~\ref{separatrix}, we examine the asymptotic matching problem close to the magnetic separatrix. We find that, due to the strong magnetic
shear in the vicinity of the separatrix, both the spacing between successive rational surfaces and the resistive layer thickness tend to zero
as the separatrix is approached. However, the rational surface spacing tends to zero faster. Consequently, there exits a radially thin region, that
straddles the magnetic separatrix, in which the resistive layers overlap. When we calculate the response to the RMP of the resistive layers in the overlap
region, as well as those in the SOL,  we find that the shielding currents excited in the layers are comparatively feeble, which implies that the plasma response to the RMP in these regions is essentially vacuum-like.  On the other hand, the shielding currents excited in resistive layers that lie in the interior of the
plasma are comparatively strong, which implies that the plasma response to the RMP in this region is essentially ideal. We are thus able to
formulate a practical, yet physically justified,  approach to calculating the response of a magnetically
diverted tokamak plasma to an RMP. In essence, we treat the region of the plasma that lies outside the inner boundary of the
overlap region, which is situated  at ${\mit\Psi}=1-\epsilon_c$, where $0<\epsilon_c \ll 1$,  as a vacuum. This means that  that rational surfaces that lie in the
region ${\mit\Psi}>1-\epsilon_c$ do not contribute to the plasma response. Hence, a plasma response calculation only needs to take a
finite number of rational surfaces into account. Our analysis enables us to determine $\epsilon_c$ from the 
edge plasma parameters. We find that $\epsilon_c=1.5\times 10^{-3} $ for an $n=1$
RMP,  and $\epsilon_c=4.8\times 10^{-3}$ for an $n=4$ RMP,  in  a typical JET H-mode plasma. 

\section*{Acknowledgements}
This research was funded by the  U.S.\ Department of Energy, Office of Science, Office of Fusion Energy Sciences under contract DE-FG02-04ER54742.

\section*{Data availability statement}
The digital data used in the figures in this paper can be obtained from the author upon reasonable request.

\end{document}